\documentclass[twocolumn]{aastex63}
\usepackage{multirow}
\usepackage{array, multirow}
\usepackage{hyperref}
\newcommand{\HI}{H{\sc i}}
\newcommand{\HII}{H{\sc ii}}
\newcommand\farcdeg{\mbox{$.\!\!^\circ$}}%
\newcommand\farcmin{\mbox{$.\mkern-4mu^\prime$}}%
\newcommand\farcsec{\mbox{$.\!\!^{\prime\prime}$}}%
\received{June 1, 2019}
\revised{January 10, 2019}
\accepted{\today}
\shorttitle{UVIT study of star forming complexes}
\shortauthors{Yadav et al.}
\graphicspath{{./}{figures/}}

\begin{document}
\title{Comparing the Inner and Outer Star Forming Complexes in the Nearby Spiral Galaxies NGC 628, NGC 5457 and NGC 6946 using UVIT Observations}

\author[0000-0002-5641-8102]{Jyoti Yadav}
\email{jyoti@iiap.res.in}

\affiliation{Indian Institute of Astrophysics, Koramangala II Block, Bangalore 560034, India}
\affiliation{Pondicherry University, R.V. Nagar, Kalapet, 605014, Puducherry, India}

\author[0000-0001-8996-6474]{Mousumi Das}

\affiliation{Indian Institute of Astrophysics, Koramangala II Block, Bangalore 560034, India}

\author[0000-0002-9981-296X]{Narendra Nath Patra}

\affiliation{Raman Research Institute, C. V. Raman Avenue, Sadashivanagar, Bengaluru 560080, India}

\author{K. S. Dwarakanath}

\affiliation{Raman Research Institute, C. V. Raman Avenue, Sadashivanagar, Bengaluru 560080, India}

\author[0000-0002-5864-7195]{P. T. Rahna}

\affiliation{CAS Key Laboratory for Research in Galaxies and Cosmology, Shanghai Astronomical Observatory, Shanghai, 200030, China}

\author[0000-0002-9762-0980]{Stacy S. McGaugh}

\affiliation{Department of Astronomy, Case Western Reserve University, Cleveland, OH 44106, USA}

\author[0000-0003-2022-1911]{James Schombert}

\affiliation{Institute for Fundamental Science, University of Oregon, Eugene, OR 97403, USA}

\author[0000-0003-4034-5137]{Jayant Murthy}

\affiliation{Indian Institute of Astrophysics, Koramangala II Block, Bangalore 560034, India}



\begin{abstract}
We present a far-UV (FUV) study of the star-forming complexes (SFCs) in three nearby galaxies using the Ultraviolet Imaging Telescope (UVIT). The galaxies are close to face-on and show significant outer disk star formation. Two of them are isolated (NGC 628, NGC 6946), and one is interacting with distant companions (NGC 5457). We compared the properties of the SFCs inside and outside the optical radius (R$_{25}$). We estimated the sizes, star formation rates (SFRs), metallicities, and the Toomre Q parameter of the SFCs. We find that the outer disk SFCs are at least ten times smaller in area than those in the inner disk. The SFR per unit area ($\Sigma_{SFR}$) in both regions have similar mean values, but the outer SFCs have a much smaller range of $\Sigma_{SFR}$. They are also metal-poor compared to the inner disk SFCs. The FUV emission is well correlated with the neutral hydrogen gas (\HI) distribution and is detected within and near several \HI~holes. Our estimation of the Q parameter in the outer disks of the two isolated galaxies suggests that their outer disks are stable (Q$>$1). However, their FUV images indicate that there is ongoing star formation in these regions. This suggests that there may be some non-luminous mass or dark matter in their outer disks, which increases the disk surface density and supports the formation of local gravitational instabilities. In the interacting galaxy, NGC 5457, the baryonic surface density is sufficient (Q$<$1) to trigger local disk instabilities in the outer disk. 

\end{abstract}




\section{Introduction} \label{sec:intro}
Star formation in the outer disks of galaxies is very different from star formation in the inner disks \citep{Bigiel2010}. This is not surprising as the inner and outer disk environments are very different, both in stellar surface densities and gas properties. The outer disks have very low stellar surface densities, and massive star-forming regions are usually not detected in these parts. They are also metal-poor, and the dust content is low, both of which are important for gas cooling and star formation. However, young stars have been detected beyond the optical radius (R$_{25}$) in H$\alpha$ emission from a few nearby galaxies by \citet{Ferguson1998}. Since then, several young star-forming complexes (SFCs) have been detected in the outer parts of both large spirals \citep{barnes.etal.2012} as well as dwarf galaxies \citep{2016Hunter}.

The other requisite for star formation is cold, neutral gas. Although the outer disks are typically rich in neutral hydrogen (\HI) gas \citep{bosma.1981}, they have very little molecular gas \citep{bicalho.etal.2019}, and CO detection from these regions is rare \citep{dessauges-zavadsky.2014}. Some studies, however, suggest that molecular hydrogen gas is not essential for star formation \citep{glover.clark.2012}. 

Although H$\alpha$ has been detected in the outer disks of a few nearby galaxies, the definitive evidence of outer disk star formation came from the discovery of extended UV (XUV) disks by the Galaxy Evolution Explorer ({\it GALEX}) mission \citep{Gil2005, Gil2007, Thilker2005, Thilker2007}. In these surveys, $\sim$ 30\% of nearby galaxies (D $<$ 40 Mpc) were found to have significant star formation at disk radii beyond R$_{25}$. Such XUV disks are found only in gas-rich galaxies and are more common in late-type spirals. A majority of them have star formation which follows the spiral structure of their inner disks; they are called type-1 XUV galaxies. A slightly smaller fraction have star formation associated with low surface brightness (LSB) outer blue disks and are called type-2 XUV galaxies. Some XUV galaxies show features common to both types and are called mixed XUV galaxies. 

Star formation in the inner disks of galaxies is mainly driven by global disk instabilities such as spiral arms \citep{rahna.etal.2018}, bars \citep{martinet.friedli.1997} or by galaxy interactions and mergers \citep{koopman.kenney.2004}. Expanding \HI~shells created by stellar winds from massive stars or triggered by supernova explosions (SNE) can also produce local star formation \citep{Richard1986}. They can result in the formation of \HI~holes of radii 10 to 1000 pc in the interstellar medium (ISM) \citep{Boomsma2008}. Star formation on both global and local scales depends on several factors, but the primary factor is the available fuel, i.e., the \HI~and molecular gas content of the disks \citep{Doyle2006}. This is evident from the tight correlation between the \HI~mass surface density and the star formation rate (SFR) surface density, also known as the Kennicutt--Schmidt law (KS law)  \citep{Kennicutt1998, bigiel.etal.2008, kennicutt.evans.2012}.  The stellar disk gravity is also important for the formation of global disk instabilities and cloud collapse. This is seen in the correlation between \HI~surface density and stellar mass densities \citep{Catinella2010, Huang2012, Maddox2015} and the correlation between the star formation rate surface density ($\Sigma_{SFR}$) and the stellar surface density \citep{shi.etal.2018}. 

The \HI~gas and star formation correlation has also been explored in different galactic environments, such as the outer parts of galaxies. It is similar in nature but has a steeper slope \citep{Bigiel2010apjl, Bigiel2010, Chayan2019}. The much lower star formation efficiencies (SFEs \citealt{Bigiel2010}) suggests that the nature of star formation in such environments may be different from the inner disks. The most likely trigger for star formation in these regions is gas accretion due to galaxy interactions or cold gas accretion from the intergalactic medium (IGM) or cosmic web \citep{lemonias.etal.2011}. Galaxies accrete gas from outside the halo virial radius over cosmic time, and the slow influx of gas fuels star formation in their outer disks. Models of cold gas accretion predict that the gas flows inwards at velocities of 20 to 60 km s$^{-1}$, from radii well beyond the edge of the stellar disks \citep{ho.etal.2019, Dekel2006, Dekel2009, Silk2012} as well as through mergers \citep{Guo2012, Kormendy2013}. 

Other examples of star formation in low-density environments are the LSB galaxies \citep{boissier.etal.2008, schombert.etal.2011}. Star formation in these galaxies is patchy, and molecular gas is nearly always absent \citep{das.etal.2006} except for some giant LSB galaxies where the disks may show more structure \citep{das.etal.2010}. The KS law has been found to be steeper in LSB galaxies \citep{wyder.etal.2009}. Their \HI-rich but low stellar density disks are similar to type-2 XUV disks, and both regions are metal-poor. The star formation in LSB dwarf galaxies and irregulars is also similar to outer disk star formation \citep{2016Hunter, patra.etal.2016}. Here again, the disks are gas-rich but metal-poor, and the stellar disks have low mass surface densities. These galaxies all lie at the lower extreme of the star formation main sequence \citep{mcgaugh.etal.2017}. Understanding star formation in the outer disks of galaxies also allows us to probe star-formation in such metal-poor, dust poor, low stellar density environments \citep{Zaritsky2007, Barnes2011}. 

If we compare the different classes of star-forming galaxies discussed in the previous paragraphs, we find that XUV galaxies are the only class of galaxies with compact SFCs in both their inner and outer disks and are also commonly found in the local universe. Furthermore, although H$\alpha$ is a widely used star formation tracer, UV emission traces both a wider class of stars and a longer star formation age. Thus, XUV galaxies are ideal laboratories to compare star formation in low density and high density environments. 
 
In this paper, we present the Ultraviolet Imaging Telescope (UVIT) far-UV (FUV) observations of star formation in the extended disk of two nearby XUV galaxies (NGC 628 and NGC 5457) and a galaxy that closely resembles a type 1 XUV galaxy (NGC 6946). The UVIT has a spatial resolution of approximately 1\farcsec5, which corresponds to a linear scale of 40 to 50 pc at distances of the galaxies in this study and is similar to the sizes of giant molecular clouds as well. This makes UVIT a good probe to detect the SFCs in the outskirts of nearby galaxies. Most of the FUV emission in galaxies is due to young O and B associations, which emit in FUV for $\sim 10^8$ years \citep{Donas1984, Schmitt2006,Thilker2007}. In general, FUV is one of the best tracers of recent star formation in galaxies \citep{kenni1998}, assuming that there is not much dust obscuration from the host galaxy. However, less massive stars also emit FUV flux in their post main sequence phase of evolution. Examples are horizontal branch stars, post asymptotic giant branch stars, and white dwarfs.

The order of this paper is as follows. In section~\ref{sec:sample_selection}, we describe the sample selection. Section~\ref{sec:observation} describes the observations and data analysis. In section~\ref{sec:results}, we present the results, and then in section~\ref{sec:discussion}, we discuss the implications of our results.

\section{Sample Selection} \label{sec:sample_selection}
This study aims to understand star formation in the extreme outer parts of galaxy disks and compare it with star formation in the inner disk regions. This includes understanding how the outer disk SFCs are related to the \HI~gas distribution. We also want to see whether the outer disk regions are locally unstable by estimating the Toomre Q parameter which requires \HI~gas mass surface densities. Hence, it was important to select galaxies that are both large and nearby so that we obtain high spatial resolution data. It is also easier to identify and study the distribution of SFCs if the galaxies are close to face-on. So we first made a sample of galaxies that lie within 10~Mpc distance, had inclination angle less than 35$^\circ$, and had prominent SFCs in their inner and outer disks.

 We also required very good \HI~data for the galaxies, both \HI~images and data cubes in order to compare the SFCs with the gas distribution and the gas velocity dispersion. The \HI~Nearby Galaxy Survey (THINGS; \citealt{Walter2008}), which has publicly available data \footnote{\url{https://www2.mpia-hd.mpg.de/THINGS/Data.html}} was a good match. We found four face-on spiral galaxies with inclinations less than 35$^\circ$ that lie within 10 Mpc distance in the THINGS data set. However, only three of them had UVIT observations. So we finally selected the following three galaxies from the initial sample: NGC 628, NGC 5457, and NGC 6946. All three galaxies have deep UVIT observations.
In the following paragraphs, we briefly discuss the properties of our sample galaxies.  

\begin{small}
\begin{deluxetable*}{lcccc ccccc c}
\tabletypesize{\footnotesize}
\tablecaption{Properties of galaxies\label{tab:galaxy_properties}}
\tablehead{
\colhead{Galaxy} & \colhead{R.A.$_{J2000}$} & \colhead{Dec.$_{J2000}$} & \colhead{Dist.} & \colhead{Incl.} &\colhead{P.A} & \colhead{Galaxy} & \colhead{Spatial Scale} & \colhead{R$_{25}$} & \colhead{Galactic} & \colhead{ log $\Sigma_{SFR}$}\\
 & (hh mm ss)& (dd mm ss) & (Mpc) & (deg) & (deg) & Type & (pc/\arcsec) & (\arcmin) & E(B-V) & (M$_{\odot}$ yr$^{-1}$ kpc$^{-2}$)
}
\decimalcolnumbers
\startdata
NGC 0628 & 01 36 41.8 & +15 47 00 & 7.3 & 7 & 20 & SA(s)c, \HII & 35.4 & 4.89 & 0.062  & $-$2.25\\
NGC 5457 & 14 03 12.6 & +54 20 57 & 7.4 & 18 & 39 & SAB(rs)cd, \HII & 35.9 & 11.99 & 0.008 & $-$2.34\\
NGC 6946 & 20 34 52.2  & +60 09 14 & 5.9 & 33 & 243 & SAB(rs)cd;Sy2;\HII & 28.6 & 5.74 & 0.3 & $-$2.92\\
\enddata
\tablecomments{Column 2 and 3 are the coordinates of galaxies in J2000.0. Column 4: Distance in Mpc. Column 5: Inclination in degrees. Column 6: Position angle in degrees. Column 7: Galaxy type as listed in NED. Column 8: Spatial Scale of galaxy. Column 9: Optical radius in arcmin. Column 10: foreground Galactic extinction taken from \citep{Schlafly2011}. Column 11: SFR surface density for NGC 628 and NGC 5457 is listed from \citet{Thilker2007} and for NGC 6946 it is taken from \citet{Leroy2008}. The R.A, Dec., Distance, Inclination, R$_{25}$ have been listed from \citet{Walter2008}.}
\end{deluxetable*}
\end{small}

\subsection{NGC 628} \label{sec:628}
NGC 628 (M 74 or the Phantom Galaxy) is a face-on disk galaxy located at a distance of 7.3~Mpc with inclination 7$^{\circ}$. It is a grand design spiral with class 9 type spiral arms \citep{Mulcahy2017, Elmegreen1984}. According to \citet{Elmegreen1984}, the arm classes 1 to 4 correspond to flocculent spiral arms, and type 5 to 12 correspond to classic grand design spiral structure.
\citet{Cornett1994} calculated the surface brightness and color profiles for this galaxy in the UV regime and concluded that the disk has significant star formation over the last 500 million years.
\citet{Zaragoza2019} analyzed the stellar population in NGC 628 using Multi Unit Spectroscopic Explorer (MUSE) optical spectra and found that the spatially resolved SFCs in NGC 628 agree with models of self-regulated star formation. The disk appears bright in the radio continuum due to the high SFR in the central part \citep{Condon1987}. 
The \HI~disk extends out to more than three times the Holmberg diameter \citep{Kamphuis1992} but is asymmetric towards the southwest part of the galaxy.

\subsection{NGC 5457}
NGC 5457 (M 101, Pinwheel Galaxy) is a face-on spiral galaxy which belongs to arm class 9 \citep{Elmegreen1984}. 
It appears asymmetrical in optical and \HI~images, which is probably due to its interaction with the satellite galaxies: NGC 5204, NGC 5474, NGC 5477, NGC 5585, and Holmberg IV \citep{Sandage1994}. The outer disk is extremely blue \citep{Mihos2013}, suggesting that there is a younger population of stars at large radii. The interaction with the companion galaxies may be the reason for the extended star formation.
The \HI~distribution and kinematics of NGC 5457 are fairly complex. \citet{Gulin1970} studied the \HI~distribution in detail and found a depletion of neutral hydrogen in the central part of this galaxy. \citet{Hulst1988} discovered high-velocity \HI~gas moving normal to the disk. 
Molecular hydrogen gas ($H_{2}$) has been detected at several locations in the outer disk \citep{Bernabe2013}. These $H_{2}$ regions are bright in FUV emission, suggesting that there is ongoing star formation in those regions.  \\

\subsection{NGC 6946}
NGC 6946 is a nearby face-on galaxy at 5.9 Mpc distance with an inclination angle of 33$^{\circ}$. Its spiral arm structure is classified as type 9 by \citet{Elmegreen1984}. 
This galaxy is also known as the Firework Galaxy as ten core-collapse supernovae have been detected in its disk since 1917 \citep{Eldridge2019}.
 It is also one of the few galaxies in the local universe that shows H$\alpha$ emission in its outer disk \citep{Ferguson1998}.
The galaxy is isolated and located in the interior of a nearby void \citep{Tully1988, Sharina1997}. However, it still shows a very large amount of star formation that extends well into its extreme outer disk. 

All the galaxies in our sample have ongoing star formation in their outer disks. However, previous UV observations had limited spatial resolution (e.g., GALEX), which was not sufficient to study the star formation process associated with the individual SFCs. Hence we carried out high-resolution FUV observations of these galaxies with the UVIT. We have also used publicly available high-resolution \HI~data to investigate the connection between \HI~gas and the SFRs. 

Table \ref{tab:galaxy_properties} summarizes the properties of the three galaxies. The observations and the data analysis are described in detail in the next section.

\section{OBSERVATIONS AND DATA ANALYSIS}\label{sec:observation}
\subsection{UV DATA}
We performed deep FUV imaging observations of the galaxy NGC 6946 using the UVIT on board the AstroSat Satellite \citep{Kumar2012}. The instrument has two co-aligned Ritchey Chretien (RC) telescopes, one for FUV (1300-1800 \AA) and another for both the NUV (2000-3000 \AA) and visible bands. The UVIT is capable of simultaneously observing in all three bands, and the visible channel is used for drift correction. It has multiple photometric filters in FUV and NUV. It has a field of view of around 28$\arcmin$ and a spatial resolution 1\farcsec4 in FUV and 1\farcsec2 in NUV, which is more than three times better than GALEX. We observed NGC 6946 and used archival UVIT data for NGC 628 and NGC 5457. 
\footnote{\url{https://astrobrowse.issdc.gov.in/astro_archive/archive/Home.jsp}}

For NGC 6946 and NGC 5457, only FUV data was available in UVIT because the NUV filter was not operational due to payload related issues. Hence, only UVIT FUV images were obtained for these two galaxies. For NGC 628, we obtained both FUV and NUV data from the UVIT archive.

\begin{table}
\centering
\caption{Details Of the UVIT observations and filters.}
\label{tab:observation}
\begin{tabular}{lc c|cr}
\toprule
 Filter & Bandpass & Zero point &  Galaxy & Exp. time\\
&  (\AA) & (mag)&  & (sec)\\
(1) & (2)  & (3) &  (4) & (5) \\
\hline
F154W & 1351-1731  & 17.778 &  NGC 0628  & 4420\\
N263M & 2495-2770 &  18.18 & NGC 0628 & 2086\\
\hline
 {F148W}  & {1231-1731} & {18.016} & NGC 5457 & 2074
 \\

& &  & NGC 6946 & 9462\\
\toprule
\end{tabular}
\end{table}

 The UVIT was able to cover the full field of NGC 6946 and NGC 628. NGC 5457 has a major diameter of 28\farcmin8, which is comparable to the UVIT field of view (28$^{\prime}$). This interacting galaxy is asymmetric in shape, and the distribution of \HI~and UV emission is lopsided.
The emission is more extended towards the northeast side of the galaxy. The GALEX FUV image showed no significant bright SFCs in the southwestern part outside R$_{25}$. Hence, the UVIT pointing during the observation was off-centered by $\sim$ 2\farcmin5~to observe the northeastern part. 

We reduced the UVIT level 1 data of NGC 6946 which was downloaded from the Indian Space Science Data Centre (ISSDC), using JUDE (Jayant's UVIT data explorer) software \citep{Jayanth2016, jayanth2017}. JUDE is a UVIT pipeline written in IDL (Interactive data language) to reduce UVIT level 1 data into scientifically useful images and photon lists. 
We used CCDLAB \citep{Joe2017} to reduce NGC 628 and NGC 5457 level 1 data. CCDLAB has a graphical user interface and produces scientific images by correcting for field distortions, flat fielding, and drift. Astrometry on each UVIT image was done by cross-matching 4 to 5 field stars with the GALEX image using the CCMAP module of IRAF. The details about the UVIT observation of each galaxy are given in Table~\ref{tab:observation}.

\begin{figure*}
    \centering
    \begin{tabular}{c c}
        \includegraphics[scale=0.5]{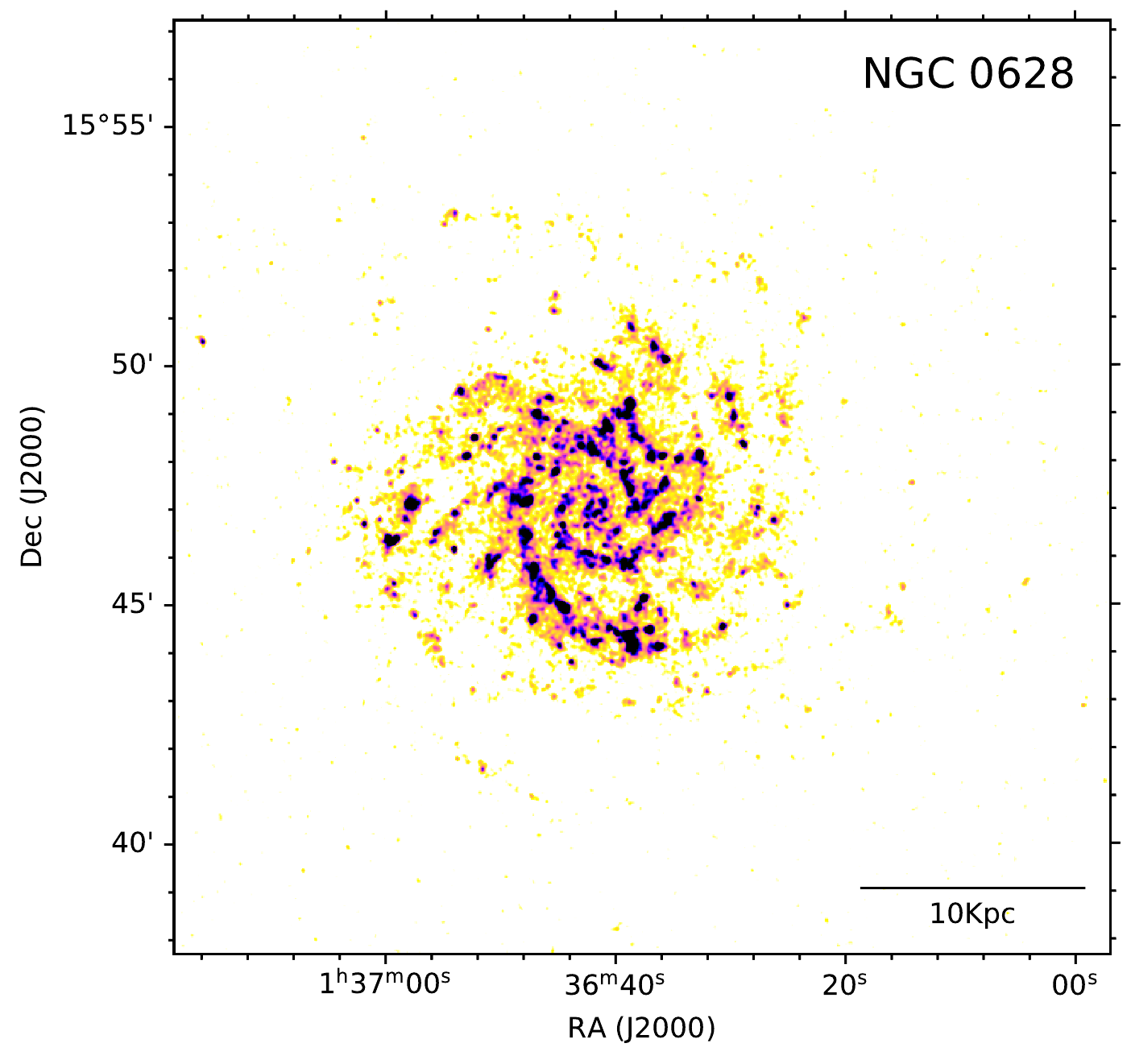} & \includegraphics[scale=0.5]{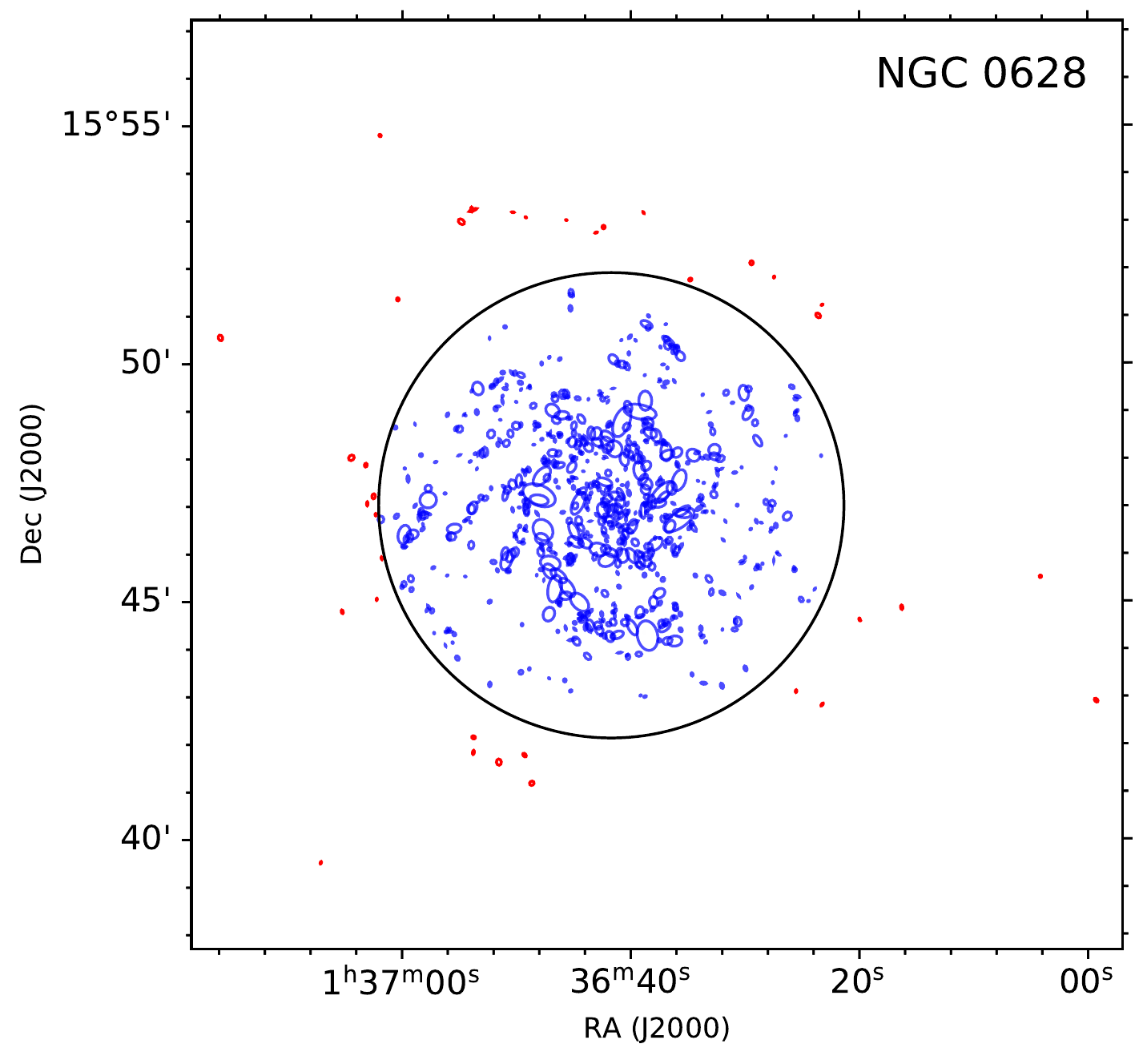}\\
        \includegraphics[scale=0.5]{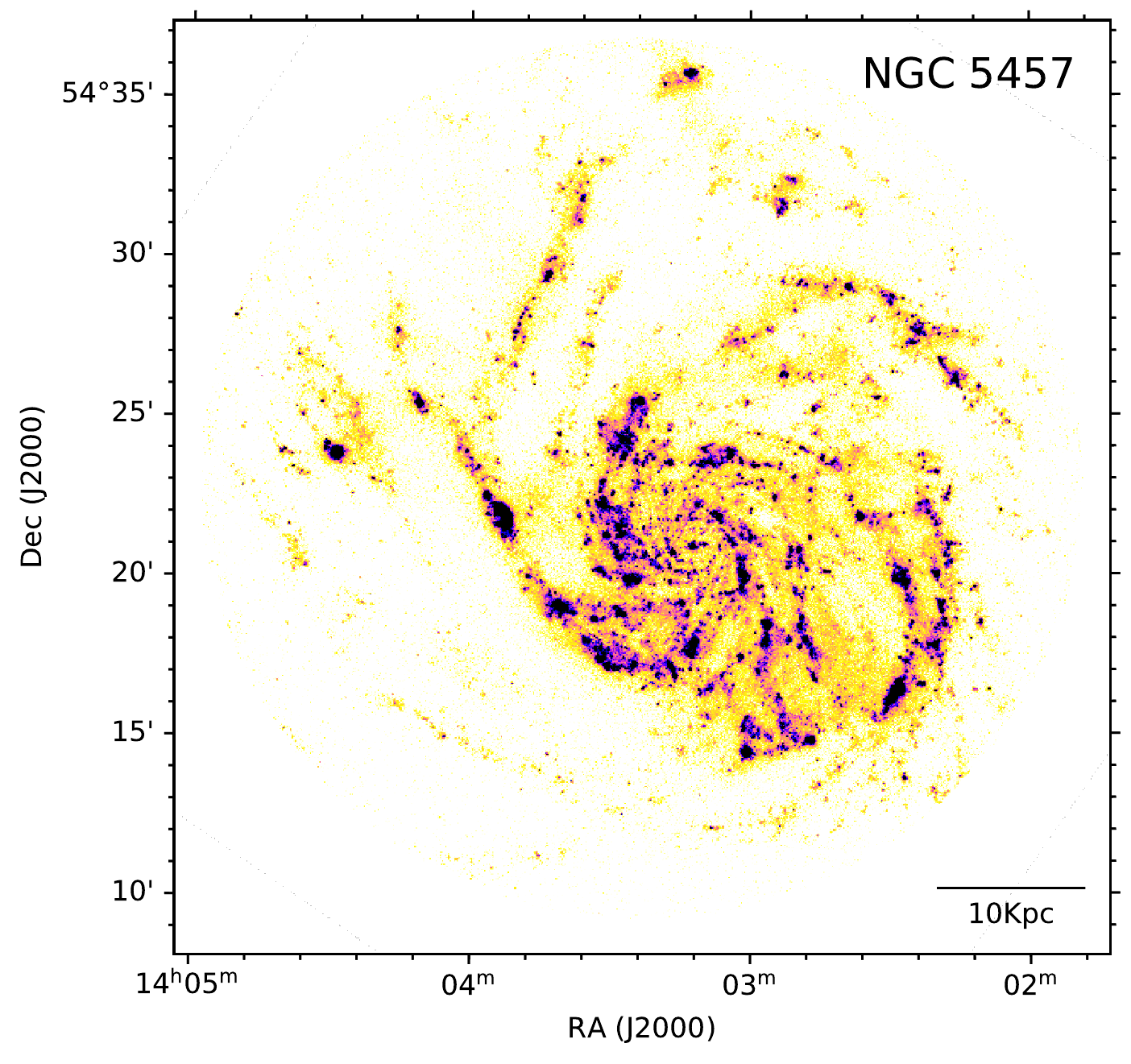} & \includegraphics[scale=0.5]{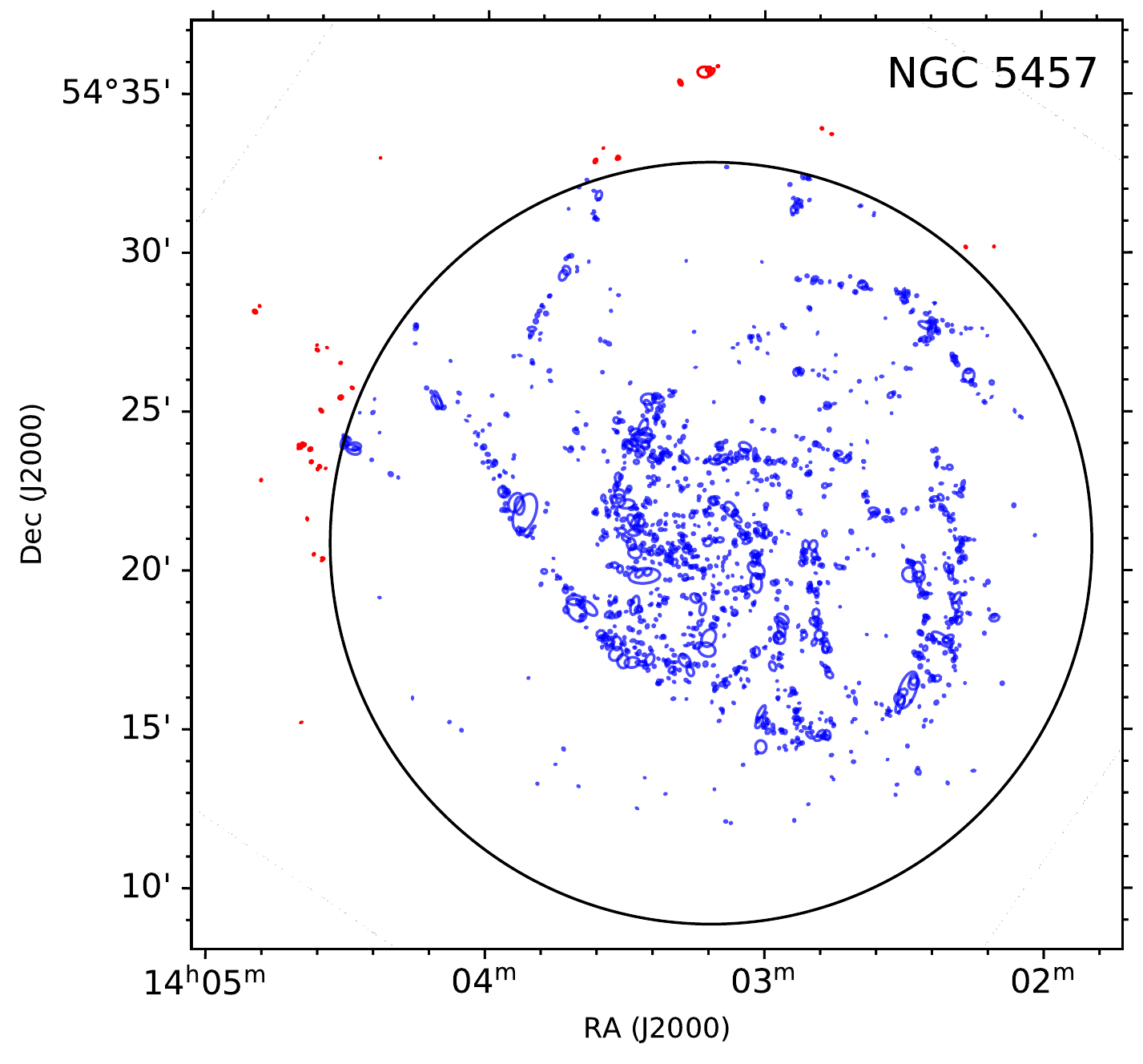}\\ 
        \includegraphics[scale=0.5]{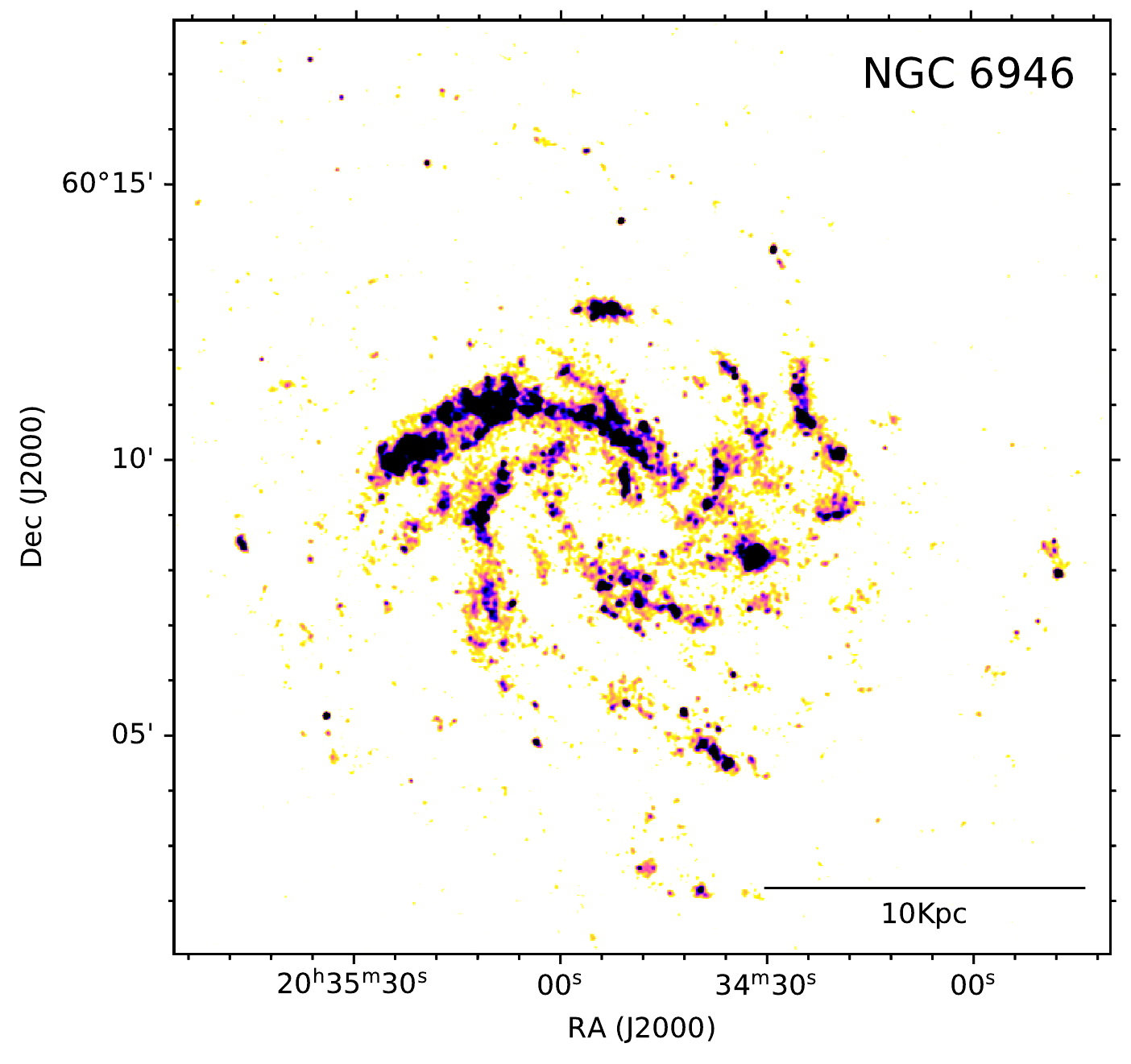} & \includegraphics[scale=0.5]{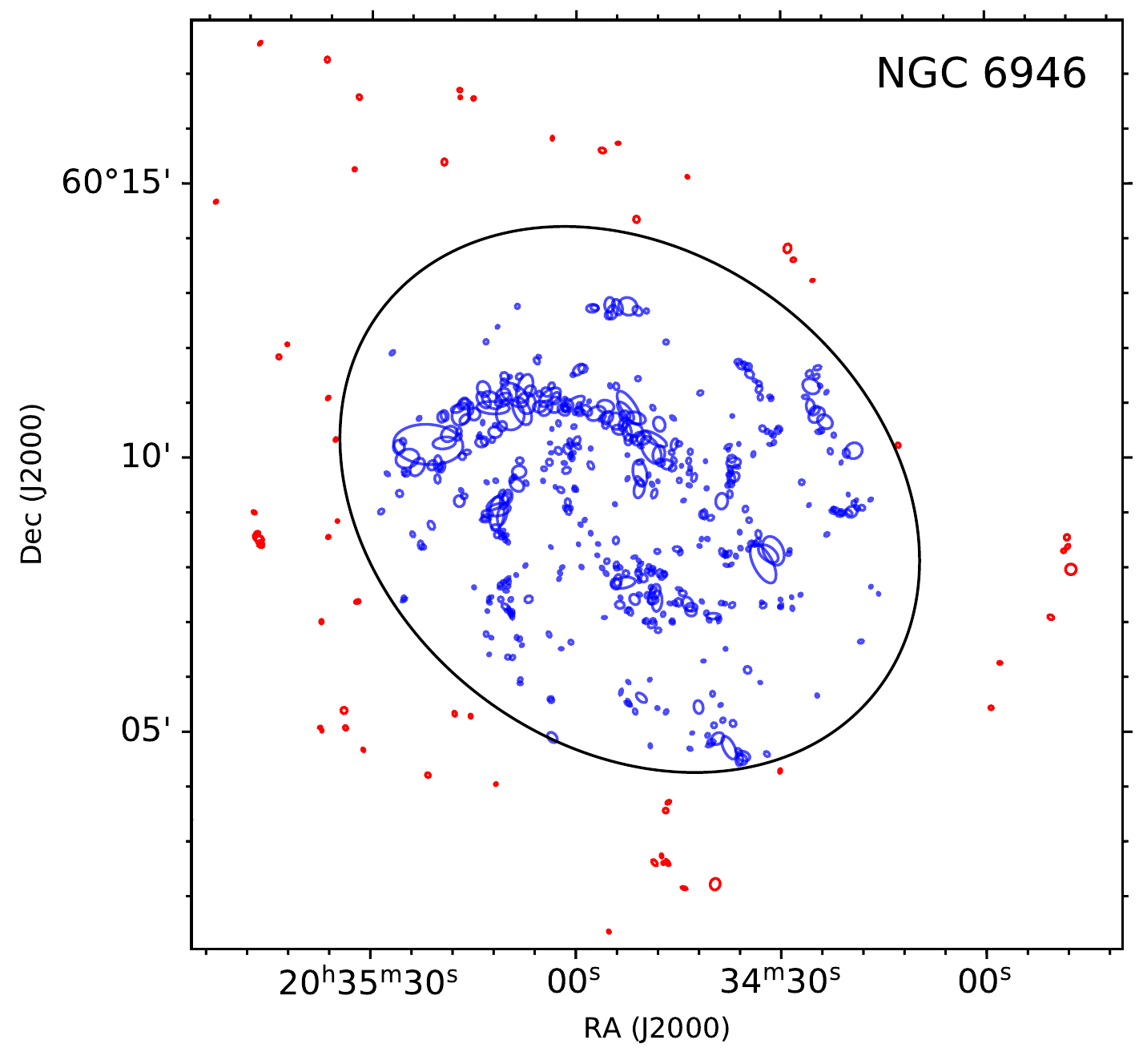}\\
    \end{tabular}
    \caption{Left panels: UVIT FUV images of NGC 628, NGC 5457 and NGC 6946. Right panels: The locations of SFCs plotted on the FUV images. The black ellipses represent the R$_{25}$ radii of the galaxies. The blue and red marks indicate the location and size of the inner and outer SFCs respectively.}
    \label{fig:uvit_images}
\end{figure*}

As mentioned earlier, the UVIT NUV data was not available for NGC 6946 and NGC 5457. Hence to constrain the metallicity of the SFCs using UV colors, which requires both FUV and NUV emission, we used GALEX images \citep{Gil2007}. GALEX has a wavelength coverage of 1344--1786 \AA{} in FUV and 1771--2831 {\AA} in NUV \citep{Morrissey2007}. The angular resolution for GALEX FUV and NUV images is 4\farcsec2 and 5\farcsec3, respectively. It has a field of view of 1\farcdeg25.

\subsection{\HI~DATA}\label{sec:HI_data}
As discussed earlier, \HI~gas plays a crucial role in sustaining long-term star formation in galaxies. Hence, it is imperative to investigate its role in the formation of SFCs in our sample galaxies, especially outside the optical radius. To do that, we have used the \HI~total intensity (MOMNT0) maps of our sample galaxies from the THINGS survey \citep{Walter2008}, which is publicly available. In the THINGS survey, 34 nearby galaxies were observed in \HI~with the Very Large Array (VLA). These observations were carried out with significantly high spatial resolutions ($\sim 6$\arcsec) and sensitivity.
For our study, it should be noted that a high spatial resolution is an essential requirement as we aim to probe the properties of SFCs of sizes of a few tens of parsecs. To date, for our sample galaxies, THINGS provides the highest quality \HI~data at high spatial resolutions. However, it should be emphasized that this data still cannot match the resolution of the FUV data ($\sim$ 1\farcsec5). Hence, for a direct comparison, we convolved the FUV images with appropriate Gaussian kernels to degrade them to the \HI~maps' resolution. A high-resolution \HI~map often suffers from a lack of sensitivity (or S/N) due to its small beam size. But for our study, a good S/N is vital to investigate the connection between gas and star formation in the SFCs. Hence, we chose to work with the MOMNT0 maps produced using the naturally weighted data cubes, which have a slightly higher S/N than what could be achieved by a robust weighted data cube \citep[see][for more details]{Walter2008}. 

\section{Results}\label{sec:results}
\subsection{Inner and Outer SFCs}\label{sec:SFCs}
\begin{figure}
    \centering
        \includegraphics[width=0.47\textwidth]{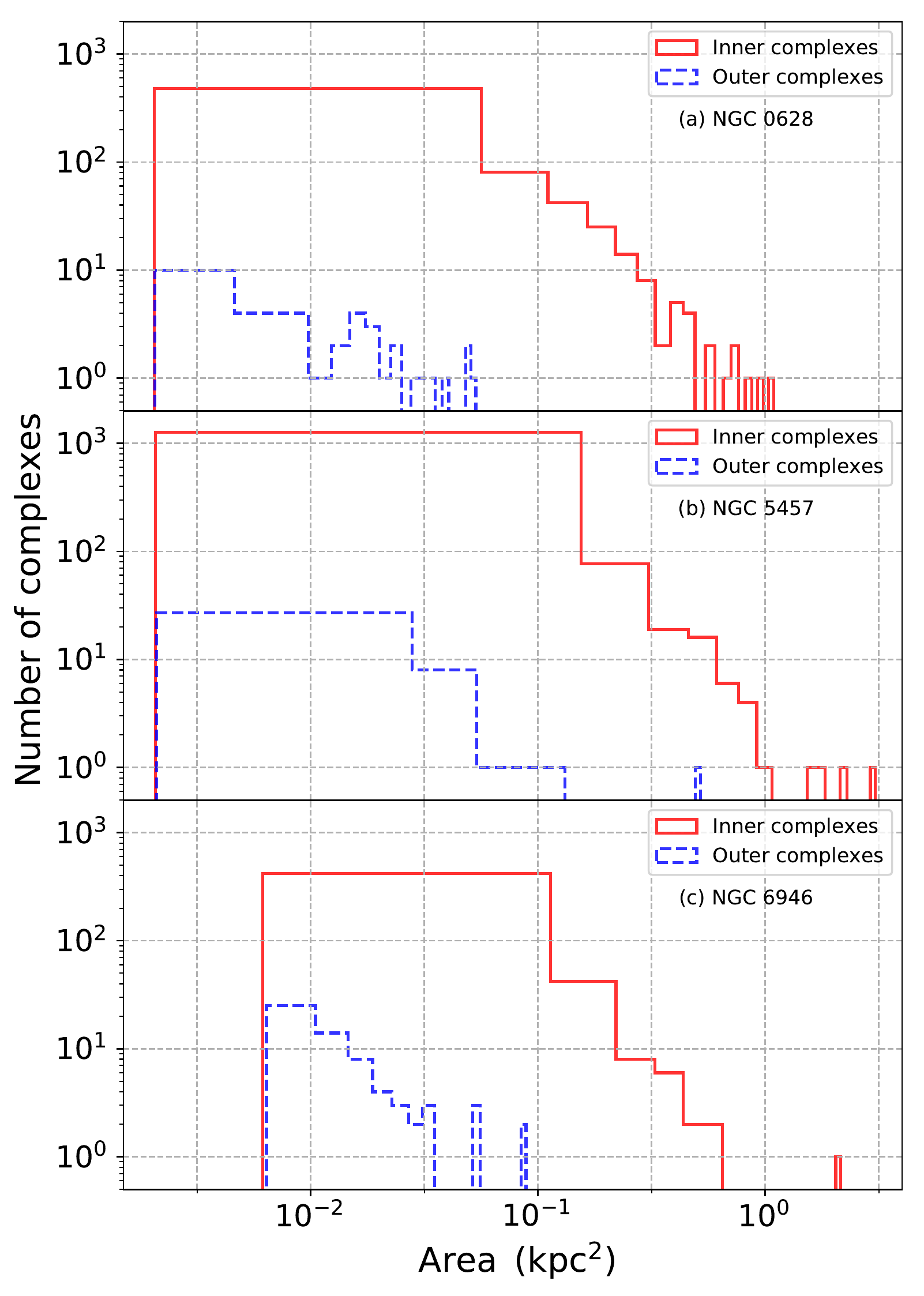}
    \caption{Histogram of area for SFCs in (a) NGC 628, (b) NGC 5457 and (c) NGC 6946. The histograms for inner and outer complexes are shown with solid and dashed lines respectively.  }
    \label{fig:area}
\end{figure}

 There is a significant amount of star formation in the outer disks of our sample galaxies, and it is clearly visible in FUV emission (Fig.~\ref{fig:uvit_images} left panels).  We first masked the foreground stars and other background sources.
We used the Python library for Source Extraction and Photometry (SExtractor, \citealt{Bertin1996}) to extract the SFCs. SExtractor is a command-line program that reads the data in FITS format and can perform multiple tasks like source detection, background estimation, and deblending. The whole process is done in several steps. First, it estimates the global statistics, and then the background is subtracted from the image. The image is then filtered and and a threshold determined. The pixels with values higher than the given threshold belong to the detected objects. Then these detected objects are deblended and cleaned. The SExtractor finds the islands which have more flux than a given threshold. For a confident and clean detection, we have used the settings recommended by \citet{Bertin1996}. We set the detection threshold to 5$\sigma$, where $\sigma$ is the global background noise, i.e., the RMS of the counts. The extracted sources are referred to as the SFCs in this work.

For NGC 628, we used the F154W (exp. time = 4420 sec) filter image to identify the SFCs because the exposure time for the broader filter F148W (1743 sec) was less. We identified the SFCs using the broad-band filter FUV F148W images of NGC 6946 and NGC 5457. We identified 706, 1425, and 544 SFCs in NGC 628, NGC 5457, and NGC 6946, respectively. 

 The spiral arms can be traced out to 4 to 5 disk scale lengths \citep{Elmegreen1998}, which is approximately equal to the R$_{25}$ radius (often called the optical disk radius). After the R$_{25}$ radius, the spiral arms become weaker. Most of the CO emission and star formation detected in galaxies lies within the optical disk. Also, the stellar disk surface density falls rapidly near the R$_{25}$ radius in both large spirals and dwarf galaxies \citep{das.etal.2020}. It is only in some galaxies, such as XUV galaxies, that there is significant UV emission beyond the optical disk. To compare the properties of star formation in the extended disk regions with the inner regions, we used the R$_{25}$ or optical radius as the demarcation between the inner and outer disk.

 The inner and outer SFCs are shown in the right side panels of  Fig.~\ref{fig:uvit_images}. Table ~\ref{tab:inner_outer_complexes} shows the number of identified inner and outer SFCs in the disks of these galaxies.

\begin{table}
\centering
\caption{Number of FUV bright SFCs inside and outside R$_{25}$. }
\label{tab:inner_outer_complexes}
\begin{tabular}{lcr} 
\toprule
 Galaxy & Inner  & Outer \\
   &      complexes & complexes  \\
(1) & (2)  & (3)  \\
\hline
 {NGC 0628}  & {668} & {38} \\
 {NGC 5457}  & {1386} & {39} \\
  {NGC 6946 }  & {480} & {64} \\
\hline
\toprule
\end{tabular}
\end{table}

The identified complexes vary in size from the inner to outer disks. We calculated the area of each complex using the formula given below 
\begin{equation}
Area  = \pi\times a\times b
\end{equation}
where a and b are the semi-major and semi-minor axes of the extracted elliptical SFCs in kpc.
Fig.~\ref{fig:area} shows the histogram of the area of these complexes for all three galaxies.

Of the three galaxies, NGC 6946 has the largest fraction (12\%) of SFCs lying beyond R$_{25}$. Whereas for NGC 628 and NGC 5457, the fractions are 6\% and 3\%, respectively. This is surprising as NGC 6946 is more isolated compared to the other two galaxies.
An important result from Fig.~\ref{fig:area} is that the outer disk SFCs are at least ten times smaller in area than the SFCs in the inner disk. So the outer disk SFCs are generally more compact than those in the inner disk. This could indicate that the star formation process may be different in the two regions, and we discuss this further in section~\ref{sec:inner_outer_discussion}.

\subsection{Estimation of Star Formation Rates}\label{sec:sfr}
We performed elliptical aperture photometry on the identified sources using Photutils, a python astropy package for photometry. We calibrated the UVIT fluxes by comparing them with GALEX images. The UVIT images were first convolved to GALEX resolution, and then a set of FUV bright isolated stars was identified in both the images. We compared the fluxes and applied the calibration factor to the UVIT images. We corrected foreground Galactic extinction using the Fitzpatrick law \citep{fitzpatrick1999} by assuming optical total to selective extinction ratio R(V)=3.1. \begin{equation}
A_{\lambda}  =R_{\lambda}\times E(B-V)
\end{equation}
where $A_{\lambda}$ is the extinction at wavelength $\lambda$, and E(B-V) is the reddening.

We calculated the $\Sigma_{SFR}$ (not corrected for internal extinction) for each complex in all three galaxies using the following formula \citep{Leroy2008, Salim2007}:

\begin{equation}
\label{eqn:sfr}
\Sigma_{SFR(UV)}  = 8.1\times10^{-2}cos(i) I_{FUV}
\end{equation}

Where $\Sigma_{SFR(UV)}$ is the SFR surface density in UV [M$_\odot$ yr$^{-1}$ kpc$^{-2}$], \textit{i} is the inclination angle of the galaxy and $I_{FUV}$ is the FUV intensity [$MJy $ ster$^{-1}$].

The error in flux ($\propto I_{FUV}$) propagates to error in $\Sigma_{SFR(UV)}$. The error in $\Sigma_{SFR(UV)}$ of each SFC is estimated as:
\begin{equation}
\label{eqn:sfr_error}
\frac{\Delta\Sigma_{SFR(UV)}}{\Sigma_{SFR(UV)}} = \frac{\Delta Flux}{Flux} = \frac{\Delta CPS}{CPS}+\frac{\Delta UC}{UC}
\end{equation}

where $\Delta\Sigma_{SFR(UV)}$ is the error in $\Sigma_{SFR(UV)}$, $\Delta$CPS is the background counts per second, CPS is the source counts per second, UC is the unit conversion factor (see eq.1 in \citealt{Tandon2017}) and $\Delta$UC is the error in unit conversion factor.
\begin{figure}
    \centering
        \includegraphics[width=0.47\textwidth]{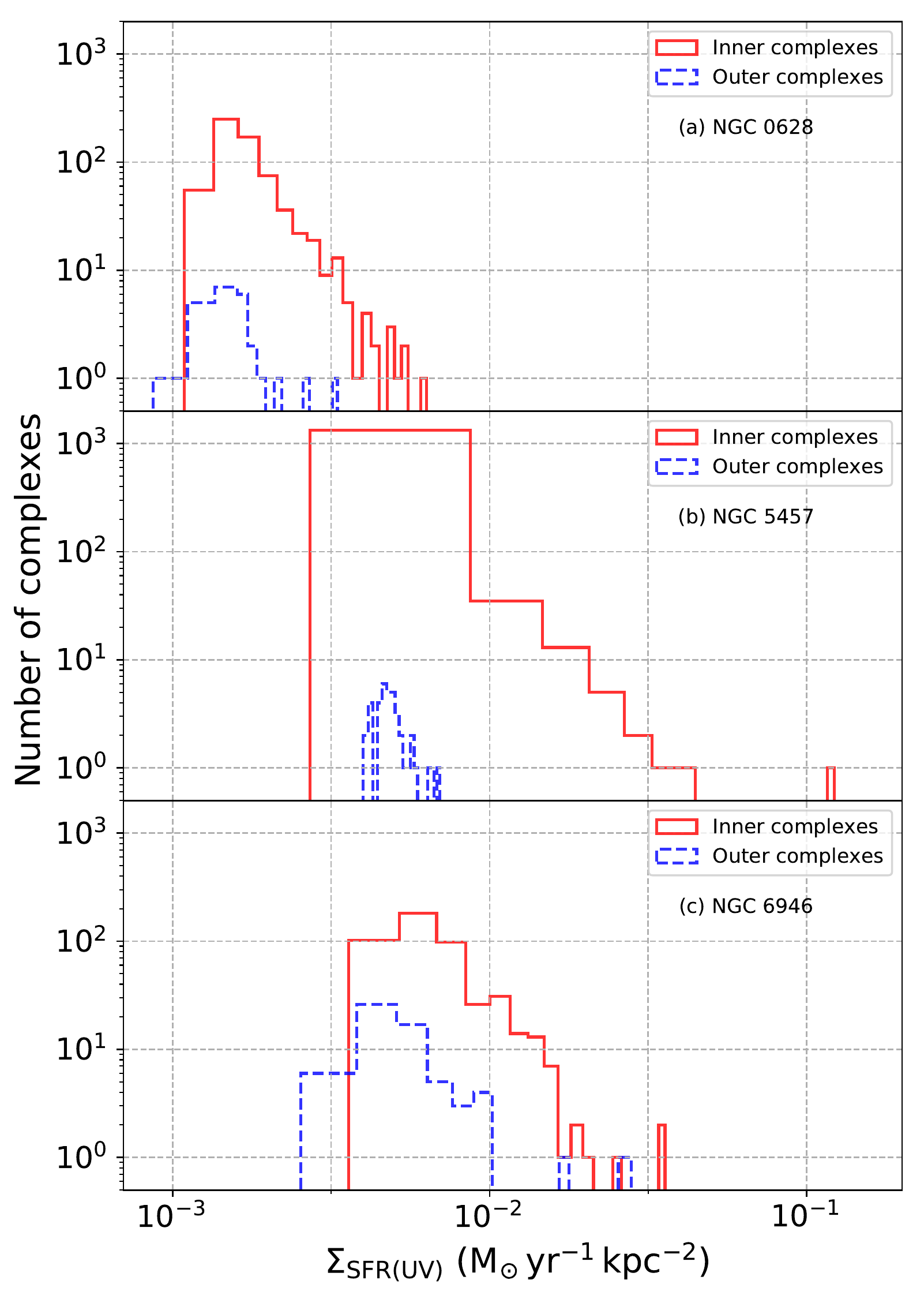}

    \caption{Histograms of the  $\Sigma_{SFR(UV)}$ in (a) NGC 628, (b) NGC 5457 and (c) NGC 6946. The histograms for inner and outer complexes are shown with solid and dashed lines respectively.  }
    \label{fig:star_frate}
\end{figure}

\begin{figure*}
    \centering
     \includegraphics[width=1\textwidth]{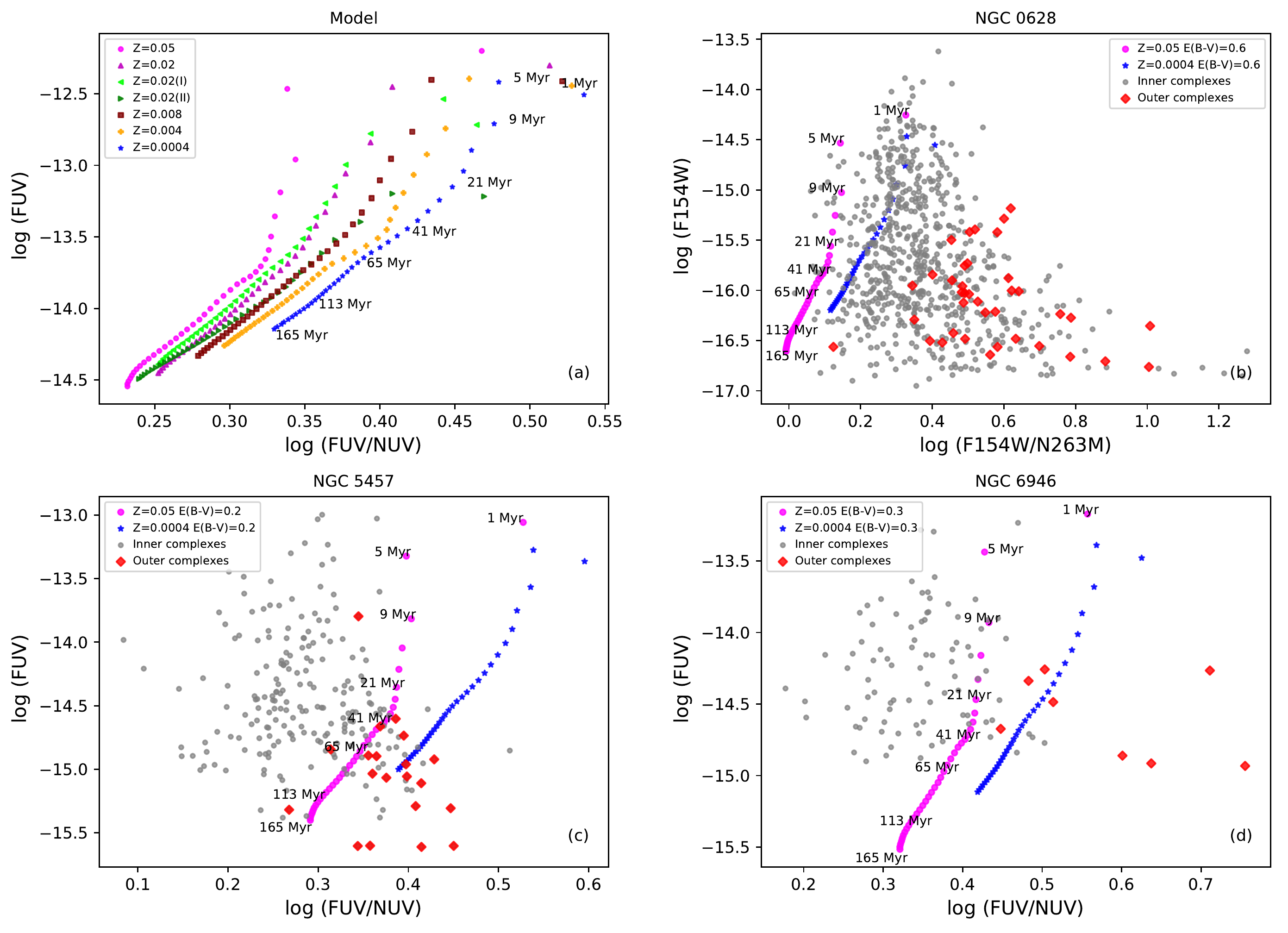}
      \caption{Log (FUV) versus log(FUV/NUV) plot for SFCs. (a) model  generated using starburst with metallicities (Z=0.0004 to Z=0.05) and ages (1 to 165Myr). Z=0.02(I) and Z=0.02(II) represents the tracks for truncated salpeter IMF and  for slope 3.3 respectively. (b), (c) and (d) panels show the SFCs in NGC 628, NGC 5457 and NGC 6946 respectively. Grey points and red diamonds represent the inner and outer SFCs respectively. The magenta (Z=0.05) and blue (Z=0.0004) tracks represents the models after including reddening of host galaxy.}
      \label{fig:age_metallicity}
\end{figure*}

Fig.~\ref{fig:star_frate} shows the histograms of the SFR surface densities ($\Sigma_{SFR(UV)}$) for the inner and outer complexes. 
For NGC 628, the maximum $\Sigma_{SFR(UV)}$ for the outer SFCs is lower than that of the inner complexes, but the lower limit of $\Sigma_{SFR(UV)}$ for the inner and outer complexes are similar. Whereas in NGC 5457, the range of $\Sigma_{SFR(UV)}$ for the outer disk is much smaller than that of the inner region. NGC 5457 has the largest range in $\Sigma_{SFR(UV)}$ for the inner SFCs.  In NGC 6946, the upper limit of  $\Sigma_{SFR(UV)}$ for the outer SFCs is not much different from that of the inner SFCs. Out of the outer 64 SFCs, two SFCs have high $\Sigma_{SFR(UV)}$ that is comparable to the $\Sigma_{SFR(UV)}$ of the inner SFCs. Of all the three galaxies, NGC 628 has the lowest $\Sigma_{SFR(UV)}$ in its outer disk. NGC 5457 and NGC 6946 have similar mean outer disk $\Sigma_{SFR(UV)}$ values.

However, it must be noted that NGC 628 and NGC 6946 have patchy dust over their disks \citep{Belley1992, Trewhella1998, Hyman2000, Cedres2012} while NGC 5457 has patchy dust in the central regions \citep{Lira2007}. Dust can absorb or scatter FUV radiation. The observed $\Sigma_{SFR(UV)}$ may thus be slightly less than the actual $\Sigma_{SFR(UV)}$ in the dusty regions; also, the actual sizes of the SFCs in the dusty region may be slightly larger than what we observe. The effect is less in the outer XUV disks since the dust content is lower in those regions \citep{ferguson.etal.1998AJ}. However, we find that the SFR for the inner complexes is higher than that for the outer SFCs (Fig.~\ref{fig:star_frate}). So our primary results will hold even after correcting for the host galaxy extinction.

\subsection{Ages and metallicities of the SFCs }\label{sec:Age}
 An accurate estimate of the metallicities and ages of the outer disk SFCs can only be determined from spectroscopic observations of the emission lines from the individual \HII~regions \citep{ferguson.etal.1998AJ,goddard.etal.2011}. However, we can constrain the ages and metallicities of the SFCs using the (FUV-NUV) colors. For NGC 6946 and NGC 5457, we have only UVIT FUV data and no NUV data. Hence, for these two galaxies, we have used GALEX FUV and NUV images. 
 The GALEX images have lower spatial resolution compared to the UVIT images, and so some of the SFCs become blended together in the GALEX images. So the number of detected SFCs in GALEX are less than those detected by UVIT. It is also more challenging to identify outer disk SFCs in the GALEX images.
However, for NGC 628, both UVIT FUV and NUV data were available, and so for this galaxy, we could measure the age and metallicity of a larger number of outer disk complexes. 

We used the Starburst99 simple stellar population (SSP, \citealt{Leitherer1999}) evolutionary models to estimate the ages of the SFCs. SSP has a detailed collection of spectrophotometric models that can constrain the properties of galaxies undergoing active star formation. We generated the synthetic stellar spectrum using starburst 99 for seven different cases. Studies show that observations of SFCs and starburst complexes are consistent with the Salpeter IMF \citep{Leitherer1998, Scalo1998}. We set the parameters by assuming instantaneous star formation and a Salpeter IMF slope = 2.35. We assumed the total stellar mass to be 10$^6$ M\textsubscript{\(\odot\)}. We considered the following range of stellar masses, with lower and upper limits given by  M$_{low}$= 10 M\textsubscript{\(\odot\)} to M$_{up}$ = 100 M\textsubscript{\(\odot\)}, and evolved the models from 10$^6$ to 2$\times$10$^8$ yr. We used the metallicities Z = 0.0004, 0.004, 0.008, 0.020 and 0.050 where Z = 0.020 equals solar metallicity.  
We also generated the spectrum for the truncated Salpeter IMF (slope = 2.35), with a mass range of 1--30 M\textsubscript{\(\odot\)} and metallicity 0.02; as well as spectra for a slope of 3.30, a stellar-mass range of 1--100 M\textsubscript{\(\odot\)} and for metallicity 0.02. Such spectra have a higher proportion of low mass stars. The models do not include nebular emission. 

Fig.~\ref{fig:age_metallicity} (a) shows the simulated data points. Metallicity decreases from left to right, and age increases from top to bottom. Fig.~\ref{fig:age_metallicity} (b), (c), (d) shows the observed data points with highest and lowest metallicity tracks. Grey and red represent the inner and outer SFCs, respectively. The outer disk points typically lie towards the right side of the lowest metallicity tracks. This indicates that outer SFCs in the three galaxies are metal-poor. Since the stars have formed in this part of the disk, the results suggest that the ISM in the outer parts of the galaxy disks is metal-poor.
FUV emission can also arise from low mass, evolved population of stars, which are generally associated with the bulges of spiral galaxies. We compared our FUV images with H$\alpha$ images in the literature. The identified star-forming complexes in FUV are clearly visible in the corresponding H$\alpha$ images as shown by \citet{Ferguson1998}, where Halpha emission is detected from the SFCs in the inner as well as outer disks of NGC 628 and NGC 6946. The emission is mostly from massive young OB stars and their associations.

\begin{figure}
    \centering
    \begin{tabular}{c}
        \includegraphics[width=0.33\textwidth]{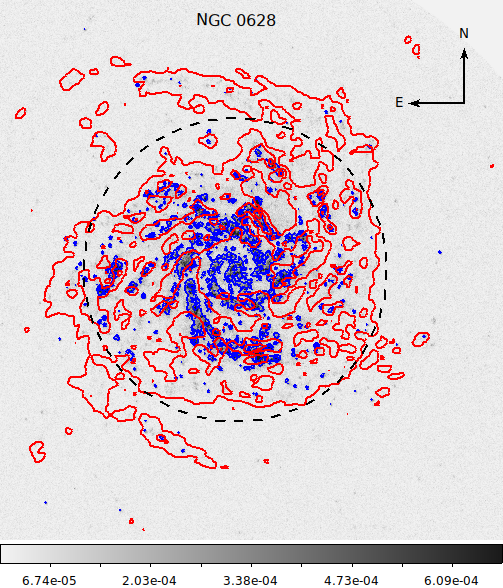} \\
        \includegraphics[width=0.33\textwidth]{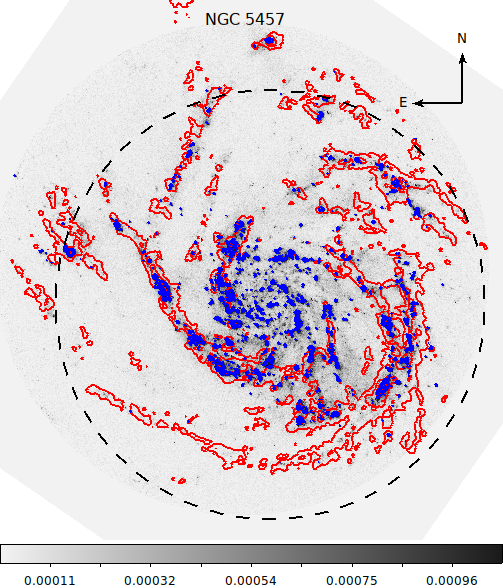} \\
        \includegraphics[width=0.33\textwidth]{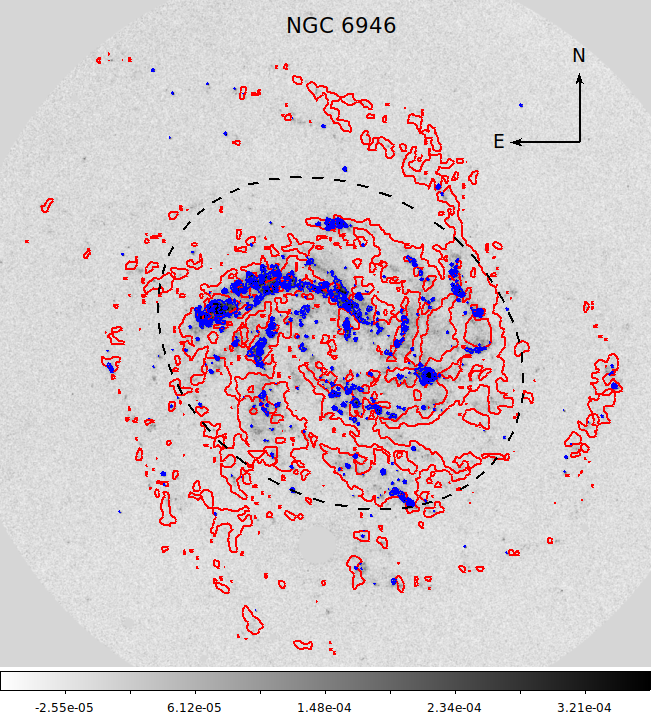} \\
    \end{tabular}
    \caption{The correlation of FUV and \HI~emission with the spiral arms. The red contours represents the \HI~emission overlaid on the background FUV image of the galaxy. The contours of FUV are represented in blue. The R$_{25}$ radius is demarcated by the black dashed aperture. }
    \label{fig:uv_HI}
\end{figure}

\begin{figure*}
    \centering
    \begin{tabular}{c }
           \includegraphics[scale=0.5]{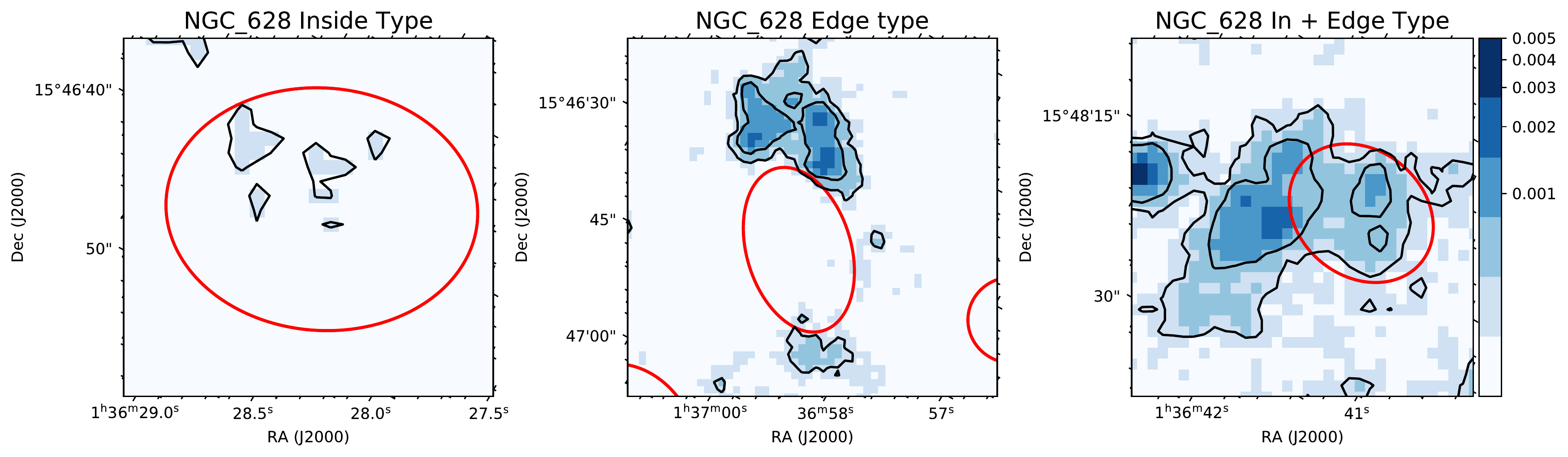}
        
        \\
        \includegraphics[scale=0.5]{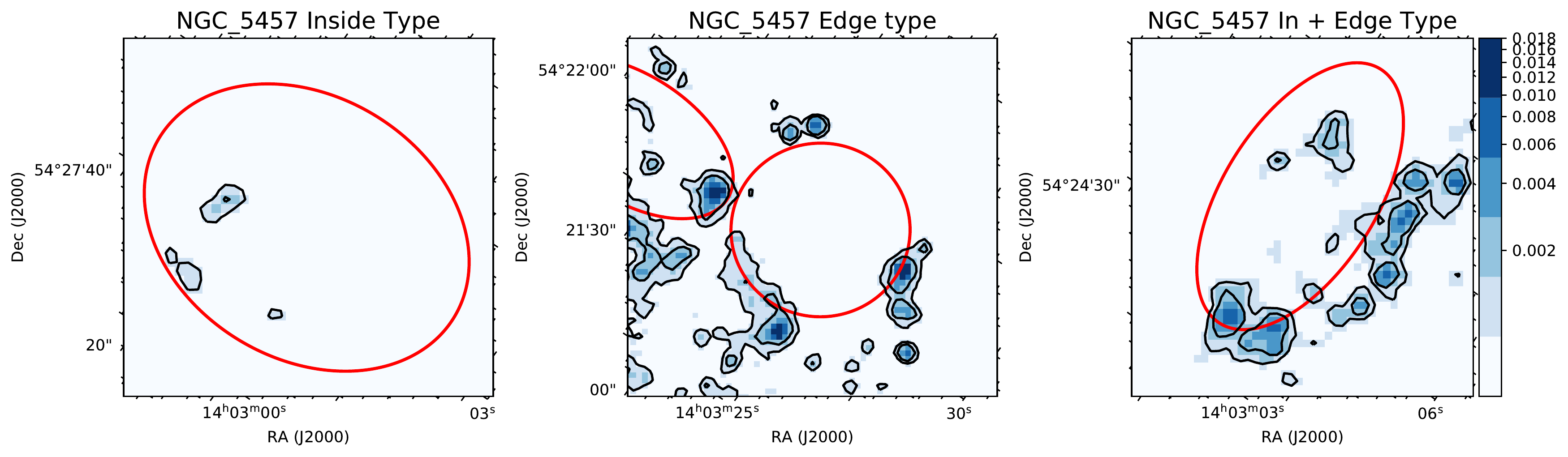}
       
         \\
        \includegraphics[scale=0.5]{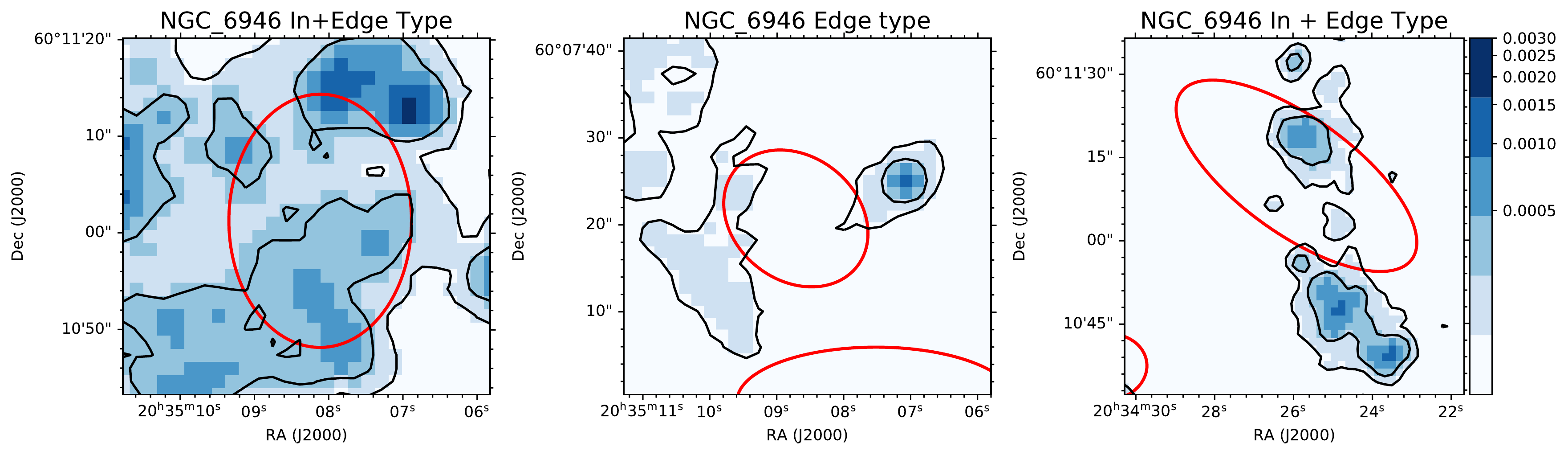}
        
    \end{tabular}
    \caption{Examples of some of the \HI~holes associated with FUV emission. The red solid ellipses  represent the \HI~hole boundaries.  The blue shaded region shown is the FUV emissions in counts per second. The FUV emission is in logarithmic scale with lower and upper bound of 2.5 and 100 sigma respectively. It is divided into 3 contour levels and 6 color-bar levels. The correlation type and name of galaxy is written on the top of each sub-panel.}
    \label{fig:Holes}
\end{figure*}

\subsection{UV correlation with \HI}\label{sec:UV_HI}
\citet{Kennicutt1998} showed that there exists a correlation between \HI~mass and the SFR, i.e., $\Sigma_{SFR}$ $\propto$ $\Sigma_{gas}^{1.4}$ with 0.2 dex scatter, and it is known as the KS law. \HI~is the primary requirement for star formation as it can cool and form molecular hydrogen in dense complexes \citep{Prochaska2009, Obreschkow2016}. The connection between \HI~and star formation has also been explored for different galactic environments, including the outer disks of galaxies \citep{Bigiel2010apjl, Bigiel2010, Chayan2019}. Hence, we expect to see an association of \HI~with star formation for our FUV images.

We compared our FUV images that trace young SFCs with the MOMNT0 images of \HI~from the THINGS Survey \citep{Walter2008}. All three galaxies show extended \HI~emission beyond their R$_{25}$ radii. Fig.~\ref{fig:uv_HI} shows the contours of \HI~($n_{HI}$ $>$10$^{21}$ cm$^{2}$) in red color overlaid on the UVIT FUV map. The blue contours trace FUV emission. The FUV complexes are clumpy and are mainly associated with the spiral arms of the galaxies, as shown in Fig.~\ref{fig:uvit_images} right panels. The \HI~contours trace the spiral arms remarkably well. Overall the FUV SFCs are very well correlated with the \HI~gas distribution, and most of the SFCs lie within the \HI~contours. The \HI~is also closely associated with the FUV emission in the outer disks of the galaxies.

\subsection{ \HI~holes and SFCs}\label{sec:HI_shells}
Supernova explosion and stellar winds can lead to the formation of kpc size holes and shells in the \HI~distribution of galaxies due to the energy injected by the associated blast waves in the ISM \citep{Tomisaka1986, McCray1987, MacLow1988}. Supernova explosions produce bubbles in the ISM, and as the bubbles expand, they sweep up mass in the ISM. If these bubbles have enough energy to tunnel through the disk gas layer, they will appear as empty holes in the \HI~maps until they are sheared out by the turbulent motion of the ISM gas and the differential rotation of the disk \citep{Boomsma2008}. Simulations also show that OB associations can inject energies of 10$^{53}$~erg into the interstellar medium, creating a hole of $\sim$ 1 kpc size \citep{Silich1996}. The radius and velocity of the expanding shells depend on the energy released and the ambient ISM parameters \citep{Bisnovatyi1995}. \citet{McCray1987} estimated that around 20\% of \HI~holes should correlate with \HII~ shells and contain ionizing O stars. Multiple studies have shown the association of star formation with \HI~shells in several galaxies, e.g., in LMC \citep{Kim1999}, SMC \citep{Stanimirovic1999}. \citet{Ehlerov2002} also studied the gravitational instability of the expanding shell and suggested that there is a critical gas surface density above which the shell becomes unstable. It can fragment and form clouds and then stars.

All three galaxies in this study have small and big holes in their corresponding \HI~maps. Using the upgraded WSRT observations, \citet{Boomsma2008} found 121 holes in NGC 6946 distributed all over the \HI~disk. \citet{2011AJBagetakos} identified 104 holes in NGC 628 and 56 holes in NGC 6946. \citet{Kamphuis1993} identified 19 holes in NGC 6946 and 52 holes in NGC 5457. We selected the holes from \citet{2011AJBagetakos} for NGC 628 and NGC 6946. For NGC 5457, we took the holes from \citet{Kamphuis1993}.

Fig.~\ref{fig:Holes} shows the correlation of some of the holes with the FUV emission. The red ellipses are the holes taken from the \HI~maps overlaid on the FUV images. We have considered the association to be real if FUV emission is above 5$\sigma$ ($\sigma$ is the global background noise) and at least two or more SFCs are associated with the hole. If FUV emission is present inside the hole, we labeled that correlation as inside type. If FUV emission is present on the edge of the hole, we labeled it as edge type, and if FUV  emission is present inside, as well as on the edge of the \HI~hole, we labeled that as (in + edge) type association. If FUV emission is absent at the edge and inside the hole, we say no correlation exists between that \HI~hole and FUV emission. 

To check whether the \HI~hole and FUV emission are true association or  chance coincidence, we created a  random distribution of holes and compared its correlation with the observations. We generated the random holes such that they have the same axes, position angle and radial distributions as that of the original holes. We randomised the location of the holes and created 10 sets of random distributions for NGC 6946. The \HI-hole and FUV emission associations were determined visually in a manner similar to that applied to the original holes. We found that a random distribution gives a correlation of 23 $\pm$ 4 \%, while the original holes indicate a 33 \% correlation. This means that the \HI-hole and FUV emission association has a significance of 2.5$\sigma$. However, \HI-holes are preferentially located in or near the spiral arms, and it is challenging to generate a random distribution of holes within the arms. So the randomized position of holes in this study gives a lower limit on the association of \HI-hole and FUV emission. It must also be noted that the \HI~ holes here are viewed in projection. In reality, they have a three-dimensional structure. Also, the massive star formation timescale, which can drive the winds producing the holes, is only a few times $10^6$ years, but FUV emission traces star formation until $\sim10^8$ years. So this value of 2.5sigma is a lower limit for the hole-UV emission correlation. Overall, from {Fig.~\ref{fig:Holes}} and the 2.5sigma excess correlation in the actual data with respect to the random data, we can confidently state that the \HI-holes and SFCs as traced by FUV, are correlated.

We find that the north--east spiral arm of NGC 6946 has most of the FUV bright SFCs inside the \HI~holes. NGC 6946 hosts one of the largest numbers of supernova events amongst all the nearby galaxies. The FUV bright complexes in NGC 5457 are also very well related to \HI~holes. The FUV emission is present on the edges of the supershells, which suggests that these SFCs may have been triggered by shell expansion.
We also separated the \HI~holes into inner and outer holes, based on whether they were within or beyond the R$_{25}$ radius. We found that there are significantly fewer holes beyond R$_{25}$ for all three galaxies. Table~\ref{tab:uv_hi_holes} shows the details about the correlation of \HI~holes with UV emission. We also found that 25\% of the holes in NGC 628, 71\% in NGC 5457, and 33\% in NGC 6946 are associated with FUV emission in our UVIT images.

\begin{table}[ht]
\caption{\HI~Holes and UV correlation}
\begin{center}
\begin{tabular}{|c|c|c|c|}
    \hline
Galaxy	&	inner/outer	&	Correlation	&	No. of holes	\\
(1) & (2) & (3) & (4) \\\hline
\multirow{6}{*}{NGC 0628 }	&	inner	&	Inside	&	1	\\
	&	(80)	&	Edge	& 15	\\
	&		&	In+ Edge	& 11	\\
	&		&	No correlation	&	53	\\  \cline{2-4}
	&	outer	&	Inside	& 0		\\
(106)	&	(26)	&	Edge	& 0	\\
	&		&	In+ Edge	& 0	\\
	&		&	No correlation	& 26		\\ \hline
\multirow{6}{*}{NGC 5457 }	&	inner	&	Inside	&	3	\\
	&	(49)	&	Edge	&	8	\\
	&		&	In+ Edge	& 23	\\	
	&		&	No correlation	&	15	\\  \cline{2-4}
	&	outer	&	Inside	& 2	\\
(52)	&	(3)	&	Edge	& 	 1	\\
	&		&	In+ Edge	& 0	\\
	&		&	No correlation	&	0	\\ \hline
\multirow{6}{*}{NGC 6946}	&	inner	&	Inside	&	0	\\
	&	(41)	&	Edge	&	9	\\
	&		&	In+ Edge	& 9	\\	
	&		&	No correlation	&	23	\\  \cline{2-4}
	&	outer	&	Inside	&	0	\\
(58)	&	(17)	&	Edge	&	0	\\
	&		&	In+ Edge	& 1	\\
	&		&	No correlation	&	16	\\ \hline	

\multicolumn{4}{m{8cm}}{Column 1: Name of the galaxy, the number in brackets represents the total number of holes in that particular galaxy. Column 2: The inner and outer number of holes that lie inside and outside R$_{25}$. Column 3: The correlation of FUV emission with the holes. The terms inner and edge indicate that the FUV emission is inside or on the edge of a hole. In+Edge indicates that FUV emission is present inside as well as on edge. No correlation indicates that there is no correlation between those holes and FUV emission. Column 4: The number of holes in that category.} \\
\end{tabular}
\end{center}
\label{tab:uv_hi_holes} 
\end{table}

\subsection{Gravitational Instability and the Q Factor in the Outer Disk}\label{sec:Q_parameter}
\begin{figure*}
\begin{center}
\begin{tabular}{c}
\resizebox{1.\textwidth}{!}{\includegraphics{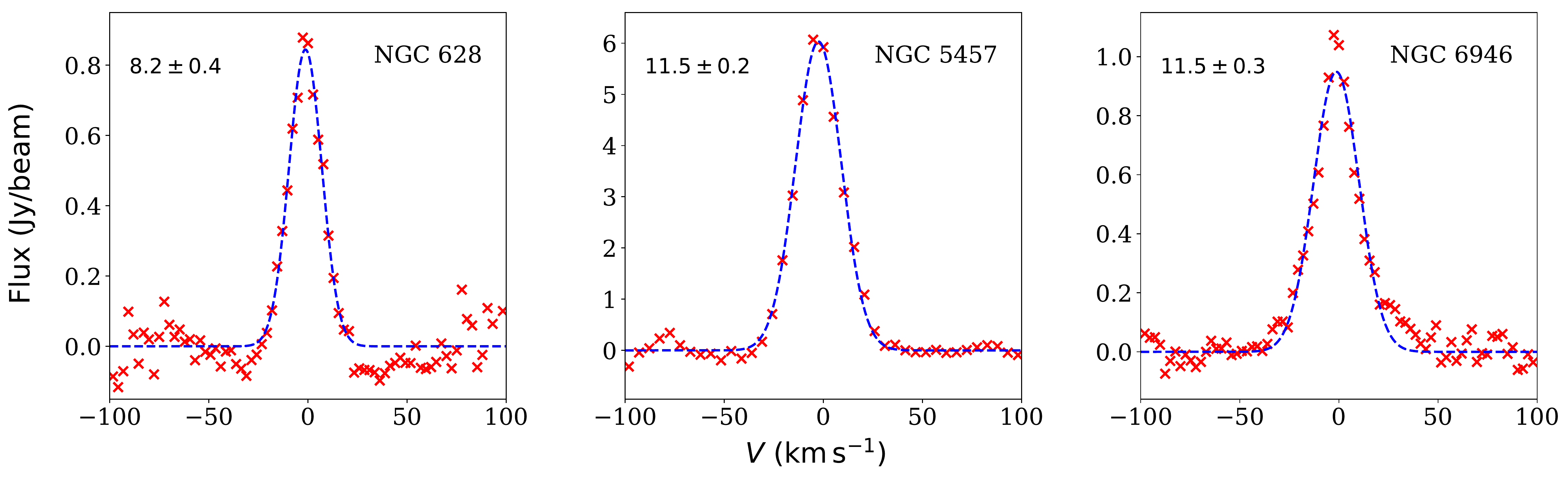}}
\end{tabular}
\end{center}
\caption{The stacked \HI~spectra of the outer SFCs in our sample galaxies. The red crosses in each panel represent the stacked spectrum, whereas the blue dashed lines represent the single-Gaussian fits to them. The respective $\sigma_g$ values are quoted in the top left corners of each panel in the units of km s$^{-1}$. See the text for more details.}
\label{stack_spec}
\end{figure*}

 In the inner disk regions, the spiral arms are the driving force that compresses the gas and triggers star formation \citep{bonnell.etal.2013}. Such triggered star formation requires strong spiral arms, which are associated with a high stellar density. Simulations have shown that the gas density in the spiral arms is three times higher than the interarm regions \citep{Hitschfeld2009}. Studies also show that massive star formation requires an ISM density of 1 g cm$^{-2}$ \citep{krumholz.mckee.2008}; and the inner disks are generally rich in molecular hydrogen gas, as well as \HI. This also means that the cloud complexes in the spiral arms are more likely to collapse compared to the interarm complexes, leading to higher star formation efficiencies (SFEs) in the spiral arms  \citep{Knapen1996}.

 In the outer disks, however, the stellar mass surface density falls rapidly, especially beyond the R$_{25}$ radius. Molecular gas is rare, and the \HI~gas has a low surface density. So this radius or the optical disk radius is a good marker to distinguish between high and low baryonic surface densities in disks (see for example \citet{das.etal.2020}). For R $>$ R$_{25}$, the spiral arm shocks are much weaker since the stellar surface density is much lower.  In such low-density environments, the primary process governing star formation must be the formation of local disk instabilities that cause the free-fall collapse of cold interstellar clouds. 

So, in general, the inner disks have triggered star formation, but the outer disks have star formation that is due to more local processes. We base this statement on the following observations derived from our study. (i) In the inner disks, the SFCs are clustered along the spiral arms. Both spiral arm shocks and strong stellar winds from the large \HII~regions can trigger star formation in the inner disks of galaxies. However, in the outer disks, the spiral arms are not so strong, and the SFCs are not strongly associated with them. They are also fairly isolated, so the winds from neighboring \HII~regions cannot trigger star formation. The lack of triggered and massive star formation is also supported by the lack of \HI~holes in the outer disks. (ii) The SFC sizes are smaller in the outer disks compared to the inner ones. (iii) There are no clear signatures of gas accretion in the outer disks in the \HI~images and velocity maps. So we cannot say for certain that gas accretion is triggering star formation in the outer disks of these galaxies. If there is gas accretion, it is happening slowly, which is expected for type~1 XUV galaxies. These differences between inner and outer disk SFCs and their location suggests that inner disk star formation is driven by global disk instabilities such as spiral arms, whereas outer disk star formation is due to local disk instabilities.

The local gravitational stability of differentially rotating disks can be investigated using the Toomre `Q' parameter \citep{binney.tremaine.2008}. It provides a quantitative estimate of the condition for local instability using parameters such as the self-gravity of the gas cloud, the thermal pressure measured by velocity dispersion in the gas, and the tidal shear of the rotating disk. The Toomre parameter \citep{toomre64} is given by
\begin{equation}
{Q}  =\frac{\kappa \sigma_{\rm g}}{\pi G\Sigma}
\label{toomre}
\end{equation}

\noindent where $\kappa$ is the epicyclic frequency, $\sigma_{\rm g}$ is the velocity dispersion, G is the gravitational constant, and $\Sigma$ is the disk mass surface density. Theoretically, a cloud is expected to be unstable under its self-gravity for $Q < 1$.
The gas surface density was calculated using the following formula
\begin{equation}
\Sigma_{HI}  = 1.4\times2.36\times10^{5}\times{D_{Mpc}}^{2}\times S_\nu \end{equation}

\begin{figure*}
    \centering
    \includegraphics[width=0.98\textwidth]{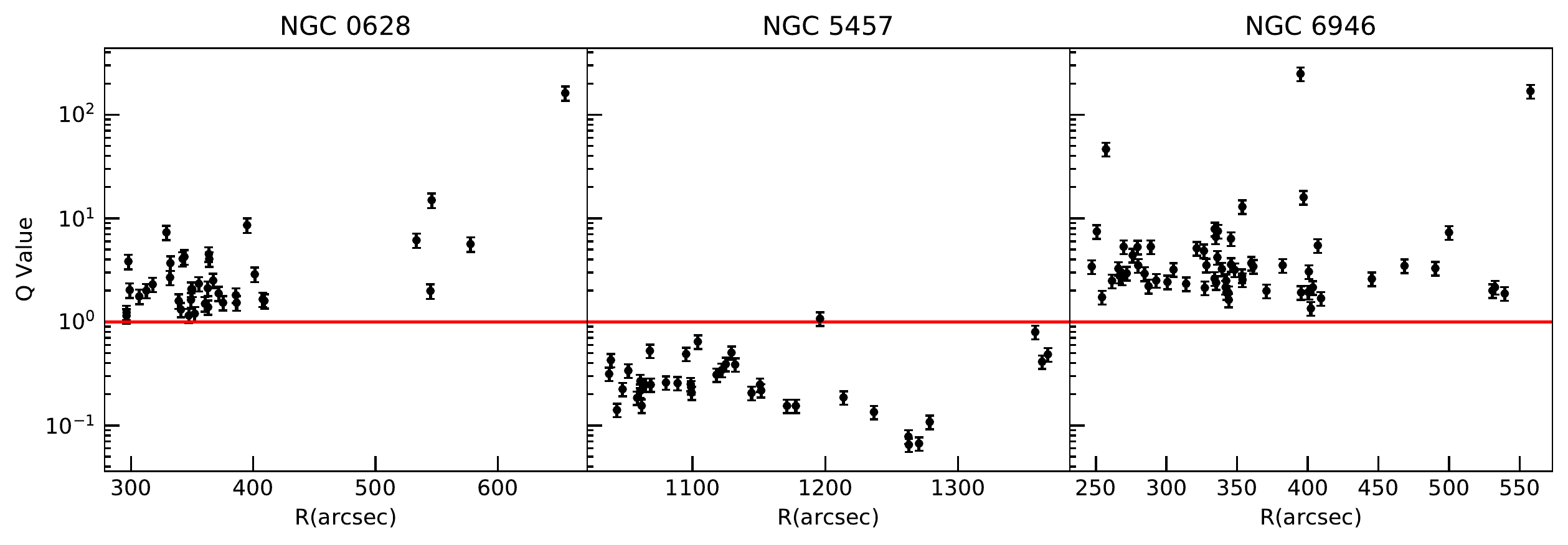}
    \caption{Toomre Q parameter for the three galaxies. The red horizontal line shows Q=1.}
    \label{fig:q_params}
\end{figure*}

 where $S_\nu$ represents the flux in Jy km s$^{-1}$. The factor 1.4 is the contribution of helium to the \HI~mass surface density. We have calculated the epicyclic frequency of individual outer SFCs using the formula $\sqrt{2}$(v/r), where v is the flat rotation curve velocity (NGC 628 \citealt{Aniyan2018}; NGC 5457 \citealt{Sofue1999}; NGC 6946 \citealt{deblok08}), and r is the deprojected radius. In doing so we have assumed that the rotation curves of the galaxies are flat in their outer disks and hence have constant velocities in the XUV regions. 

We are primarily interested in studying  the stability of the disk in the vicinity of the SFCs beyond the R$_{25}$ radius. In these regions, the mass component can be considered to be the \HI~gas as we do not observe any detectable emission from the stellar disk. Also, no considerable amount of molecular gas is detected at these outer radii \citep[see, e.g.,][]{schruba11}. Due to these reasons, we adopt the \HI~surface density and \HI~velocity dispersion ($\sigma_{\rm g}$) for determining Q. 

To estimate the \HI~surface density near the star-forming complexes, we used the MOMNT0 maps of our sample galaxies from the THINGS survey (see, \S~\ref{sec:HI_data} for more details). We identified the regions near the SFCs in the \HI~maps, using their geometrical parameters, as shown in table~\ref{tab:FUV_sfr_radec}. We used the total \HI~intensity values of all the pixels within an SFC to compute the average face-on (corrected for inclination) \HI~surface density ($ \Sigma $) for that particular SFC. 

However, estimating $\sigma_{\rm g}$ for individual complexes is not straightforward. The width of the emission spectra from a region is generally used as the proxy of the gas velocity dispersion. For quantitative estimation, an emission spectrum is fitted with a Gaussian function. The sigma of this fitted Gaussian function can be used as a measure of $\sigma_{\rm g}$. However, due to the small size, the strength of \HI~emission originating from individual complexes would be very low. As a result, the S/N of the \HI~spectrum from an SFC (typically $\lesssim 5$ on the outskirts, see, e.g., \citealt{ianjamasimanana12}) is not good enough to comfortably fit it with a single Gaussian to obtain $\sigma_{\rm g}$. Furthermore, it is well known that at low S/N, the width of the \HI~spectral line can considerably underestimate the actual $\sigma_{\rm g}$ \citep{patra18c,patra20a}. At low S/N, an \HI~spectrum only trace the peak of the emission, reducing the measurable width artificially. To overcome this problem, we adopt a similar method as used by \citet{ianjamasimanana12} \citep[see also,][]{patra20b,saponara20} and employ a spectral stacking technique to determine $\sigma_{\rm g}$ for the SFCs in the galaxies. Assuming that the \HI~spectra from all the SFCs in the outer disks are similar, we stacked them together after aligning their central velocities. To estimate the central velocity of an \HI~spectral line, we used a method similar to that used by \citet{deblok08}. We fitted the \HI~spectra with Gaussian-Hermite polynomials of order three to determine their central velocities. All the spectra from different SFCs are then aligned to their central velocities and stacked to produce a single high-S/N spectrum. This stacked spectrum can then be considered a representative of the \HI~profiles of the SFCs in the outer parts of the galaxies. We note that a considerable variation in $\sigma_g$ could be possible for different SFCs; however, the stacked spectra are expected to provide an average estimate of $\sigma_g$ in the outer disk. 

In Fig.~\ref{stack_spec}, we plot the respective stacked spectra (red crosses) for the galaxies. As shown from the figure, the S/N of the stacked spectra are high enough to estimate the $\sigma_g$ with a reasonable degree of confidence. We further fitted these spectra with single-Gaussian profiles (blue dashed lines) to determine the $\sigma_g$. Ideally, double-Gaussian fits describe the spectra better for a two-component ISM; however, as we are interested in the overall stability of the gas layer against gravity, we chose to fit the spectra with single-Gaussian profiles. Corresponding $\sigma_g$ values for each galaxy are quoted at the top left of the respective panels in Fig.~\ref{stack_spec} in the units of $\rm km \thinspace s^{-1}$.

 We note here that the resolution of the \HI~map from the THINGS survey is $\sim 6\arcsec$ (robust weighted), which is larger than the spatial resolution of the UVIT data ($\sim 1.5\arcsec$). For consistency, we tried convolving the UVIT data to the resolution of the \HI~data and identifying the SFCs. However, at this poor resolution, we detect $\textless 15\%$ of the SFCs detected in the original resolution. Hence, for calculating, $\sigma_{\rm g}$, we choose to use the location of the SFCs as detected in the original UVIT resolution. Our estimated $\sigma_{\rm g}$ values are expected to represent an average velocity dispersion around the SFCs. However, we also emphasize that the change in $\sigma_{\rm g}$ within $\approx 6\arcsec$ beam ($\sim 200$ pc at distances of the galaxies in this study) would be minimal as the \HI~clouds form structures at similar scales ($\sim 150$ pc, see, e.g., \citet[][]{braun.1997}).

The Q parameter calculated for the outer complexes are listed in Table~\ref{table:qvalue} and shown in Fig.~\ref{fig:q_params}. All the outer complexes in NGC 6946 and NGC 628 have Q $>$ 1, but in the case of NGC 5457, 97 \% of the outer complexes have Q $<$ 1. Thus, in terms of the Q parameter, the outer disks of NGC 628 and NGC 6946 appear to be more stable than NGC 5457. The difference between NGC 5457 and the other two galaxies is that NGC 5457 is interacting with nearby companions. The tidal interaction results in extended spiral arms that make the disk more unstable via cloud collisions and shocks and results in the formation of dense \HI~gas.

For the other two galaxies, NGC 628 and NGC 6946, the disk stability as indicated by Q $>$ 1, may be misleading as there is clearly significant star formation happening in the outer disks of these galaxies. This has important implications for the disk mass surface density, as discussed in the next section.

\section{Discussion}\label{sec:discussion}
\subsection{ The difference between star formation in the inner and outer disks of galaxies}\label{sec:inner_outer_discussion}
One of the main results of our study is that star formation in the inner and outer disk regions are very different from each other. There are clearly fewer SFCs in the outer disk than the inner region, and there is less star formation in the outer parts of the galaxies. This is not surprising as outer disk regions are adverse environments for star formation as both the stellar and \HI~gas surface densities are low \citep{das.etal.2020}.

As mentioned in section \ref{sec:SFCs}, the outer disk SFCs are also smaller in size and more compact than those in the inner regions, and Fig.~\ref{fig:area} shows this quite clearly. This indicates that the star formation process is different for the two regions. As discussed in section \ref{sec:Q_parameter}, star formation in the inner disk is triggered by the spiral arms. The stellar and \HI~disk surface densities in the inner disk are higher, and as a result, the SFCs are larger, less compact, and more massive than those in the outer disk. The outer disk star formation is due to local disk instabilities, which could be stochastic in nature or perhaps driven by cold gas accretion. Thus, the difference in SFC sizes in the two regions within a galaxy clearly shows that the inner disk regions host triggered star formation, whereas the outer disk regions have star formation due to local disk instabilities. 

Another factor that plays an important role is the metallicity of the inner and outer disk regions. The central disks of the galaxies are more metal-rich than the outer parts (Fig.~\ref{fig:age_metallicity}). The metals can cool the gas efficiently, and this helps in the formation of larger complexes. However, the outer disks have lower metallicities, and hence the cooling is less efficient, leading to smaller disk instabilities and hence smaller SFCs.

The FUV emission and hence SFRs of the SFCs will be affected by the  patchy dust extinction in the galaxies and the effect will be fairly random over the  SFCs. But the dust extinction in the inner disk is higher, and thus some SFCs may not be visible in FUV due to local dust extinction. So the observed star formation rates that we derived from FUV will then be less than the actual star formation rate in the dusty regions. However, we see that star formation rates for the inner complexes are higher than the outer SFCs, and so the results of our comparison of inner and outer disk star formation will hold even after correcting for the host galaxy extinction. Also, dust can scatter or absorb FUV radiation, so the actual sizes of the SFCs in the dusty region may be slightly larger. The metallicity of the SFCs in the patchy regions will change, but the overall conclusion that the outer SFCs are metal-poor will remain valid.

\subsection{Comparing the star formation in interacting and isolated galaxies}
\citet{Young1986} studied a sample of isolated and interacting galaxies and found that interacting galaxies have higher star formation efficiencies. The star formation during galaxy interactions depends on the environment in which the galaxies reside and especially depends on the mass, gas content, and distance of the companion. However, \citet{Pearson2019} suggested that the SFR of merging galaxies is not significantly different from non-merging galaxies. In our study, the histograms (Fig.~\ref{fig:star_frate}) show that the $\Sigma_{SFR(UV)}$ in the interacting galaxy (NGC 5457) is not significantly different from the isolated galaxies (NGC 628 and NGC 6946). 
However, it must be noted that the companion galaxies of NGC 5457 are relatively smaller in size and have low gas content. The interaction of NGC 5457 is a minor interaction, and this may be why it has not produced any significant effect on enhancing the $\Sigma_{SFR}$. 

In our study, we also note that there are fewer SFCs outside R$_{25}$ in the interacting galaxy (NGC 5457) as compared to the isolated galaxies (NGC 628 and NGC 6946, Table ~\ref{tab:inner_outer_complexes}). Although this seems surprising at first, it indicates that other factors such as star formation quenching due to gas stripping (by ram pressure or tidal forces) or the strength of the tidal shocks may be important for determining the SFRs in the outer disk which is more prone to interaction.
 
 \subsection{The difference in stellar metallicities of the inner and outer disks} \label{sec:metallicity}
 The outer SFCs in all three galaxies are metal-poor (Fig.~\ref{fig:age_metallicity}), and this suggests that the ISM is also metal-poor. \citet{Hwang2019} studied a sample of 1222 late-type star-forming galaxies and found that the gas phase metallicity in the outer disk is significantly lower than that expected from stellar mass surface density and metallicity correlations. They suggested that the presence of such metal-poor gas in isolated galaxies is linked to the accretion of gas from mergers, interactions, and from the intergalactic medium since a fresh supply of gas is required for continuous star formation.   
 
Another reason why the outer disk ISM is metal-poor is that metals are produced in massive stars via nucleosynthesis. Stellar winds and supernova explosions enrich the ISM with metals. Since the fraction of \HI~holes is low in the outer disks compared to the inner disks, there have been fewer supernova events and generally less massive star formation in the outer disks. This is also supported by our results, as we do not see as many large SFCs in the outer disks of our galaxies, which indicates there is not enough metal enrichment. Some of the metal-poor gas maybe pristine gas accreted from the IGM or cosmic web.

We used GALEX data for two of the sample galaxies to compare the metallicities of the SFCs in the inner and outer regions (NGC 5457 and NGC 6946) and UVIT images for one galaxy (NGC 628). The GALEX FUV images have a full width half maximum (FWHM) of $\sim5^{\prime\prime}$. So we can only detect the larger SFCs in the GALEX images.  Hence, we did not detect as many outer disk SFCs in the GALEX images as compared to the UVIT images. However, the general trend of lower metallicity in the outer disk holds in both UVIT and GALEX observations.

 \subsection{The \HI~holes and the associated FUV emission}
The \HI~gas has a higher density along the spiral arms in the inner disk region (Fig.~\ref{fig:uv_HI}).
All three galaxies that we studied have prominent \HI~holes, and some of them are a few kpc in size (Fig.~\ref{fig:Holes}). These larger \HI~holes are likely to be due to supernova explosions that lead to the formation of shocks that compress the gas in the surrounding ISM. When radiative cooling becomes dominant in the gas around the holes, the cooler gas can collapse to form molecular clouds that support star formation. Hence, star formation is often detected at the edges of the \HI~holes. Not surprisingly, the molecular gas complexes are often found at the edges of \HI~holes/shells, suggesting that they may have formed due to compressed, cooled gas or expanding shells \citep{Dawson2011, Dawson2013}.

The ISM density normal to the disk is less than that in the plane of the disk, so the \HI~holes can expand in the vertical direction more rapidly than in the horizontal direction. The material that the hole is pushing is carried along with it to some height from the disk plane. However, the \HI~holes in which the FUV emission is detected inside the hole suggests that the hole is in the early stages of evolution. 
The FUV emission inside a hole could be due to the massive stars which formed in a cluster and have not moved significantly from their point of origin. During their evolution, they produce intense radiation, ionizing the ISM and creating the \HI~hole. Thus, tracing the correlation of FUV emission with the \HI~hole structures can help us understand how stellar winds and supernova shells interact with the ISM in galaxies. We have not explored this in much detail in this paper, as it is beyond the scope of the present study. However, it is clear from Fig.~\ref{fig:Holes} that the FUV-\HI~hole correlation is important for understanding ISM evolution.
 
 \subsection{Outer disk star formation and its implications for disk dark matter }
Since local instabilities are the main cause of star formation in the outer disks of galaxies, the Toomre Q parameter can be used to check the gravitational stability of these regions \citep{das.etal.2019}. We find that the Q parameter in the outer parts of the isolated galaxies NGC 628 and NGC 6946 is greater than 1, suggesting that the gas is stable in these regions (Fig.~\ref{fig:q_params}). Since the Q value is generally expected to be less than one for star-forming regions, the $Q>1$ values indicate that the self-gravity of the \HI~gas is not sufficient to make the outer regions unstable. However, there is evidently ongoing star formation in these regions. This could either because the disk stability does not happen at exactly $Q=1$ or there is additional disk mass, apart from \HI, contributing to the total disk mass surface density, $\Sigma$.
    
One of the ways in which $\Sigma$ could increase is by including the mass of molecular hydrogen near the SFCs. However, molecular hydrogen is not generally detected in the outer disks of galaxies, and so if it is present, its mass is probably quite low. Another possibility could be that there is dark matter in the disks, which helps in the vertical support of the \HI~gas disk \citep{das.etal.2020} and increases the $\Sigma$ so that local instabilities can form \citep{das.etal.2019}. The disk dark matter could be due to an oblate dark matter halo, in which the inner flattened part of the halo is associated with the disk \citep{olling.1996}. Studies suggest that different forms of dark matter may exist (e.g., self-interacting dark matter) and may settle into a disk \citep{fan.etal.2013}. The disk dark matter could have also accumulated within the outer disks from the accretion of smaller satellite galaxies \citep{bonaca.etal.2019}. During such minor mergers, the dark matter may be stripped from the halos of the smaller galaxies and remain in the outer disks. The existence of Q $>$ 1 in the outer star-forming disks of galaxies may be a powerful means to constrain the disk dark matter surface density $\Sigma(DM)$ in galaxies, where $\Sigma = \Sigma(H\textsc{i}) + \Sigma(DM)$. 

In the case of NGC 5457, which is interacting at a distance with the dwarf galaxy NGC 5474,  Q $<$ 1 in the outer disk (Fig.~\ref{fig:q_params}), which suggests that the gravity of the \HI~gas is sufficient to make those complexes unstable. Isolated galaxies have dark matter halos that can extend out to a radius of a few 100 kpc. In contrast, in interacting galaxies, the encounter may tidally perturb or even strip off the dark matter halo from the outer part. Also, during interactions, the tidal forces between galaxies may compress the gas, enhance cloud collisions, and lead to star formation. Hence, disk dark matter need not play a role in supporting star formation in interacting galaxies but may be important for isolated galaxies. 

\section{Conclusions}\label{sec:conclusion}
In this study, we have compared the properties of the inner and outer SFCs in three nearby face-on spiral galaxies with FUV observations. The observations have been done using the UVIT, which has a good enough spatial resolution to detect a large number of SFCs in the inner and outer regions.\\
{\bf 1.}~The SFCs outside the R$_{25}$ radius have a smaller size and are more compact than the SFCs, which lie in the inner disk. This implies that the SFCs in the outer disk regions are formed from local disk instabilities, whereas those in the inner disk are formed due to global disk instabilities, which in these galaxies is the spiral arms.\\
{\bf 2.}~The surface densities of the star formation rate, $\Sigma_{SFR(UV)}$ in the outer disks of the isolated galaxies NGC 628 and NGC 6946 have surprisingly similar values to those in the inner regions.  However, in the interacting galaxy NGC 5457, the outer disk $\Sigma_{SFR(UV)}$ has a much narrower range compared to that in the inner disk. \\
{\bf 3.}~The mean $\Sigma_{SFR(UV)}$ in the inner disk of the interacting galaxy NGC 5457 is similar to that in the isolated galaxies NGC 6946 and NGC 628.\\
{\bf 4.}~ We find that the SFCs in the outer disks are metal-poor compared to the inner disk SFCs. This means that the environment and the ISM gas is also metal-poor in the outer disk. We were able to compare the metallicity for a larger number of SFCs in NGC 628 for which we have both FUV and NUV data from UVIT.\\
{\bf 5.}~ Our study shows that the FUV emission is well correlated with the \HI~gas. We have detected FUV emission within and near several \HI~holes in all three galaxies. However, most of the \HI~holes were found to lie within the R$_{25}$ radius. This suggests that massive star formation, which often results in the formation of \HI~holes, is not common in the outer disks of galaxies. Also, expanding shells and supernova explosions that can trigger star formation around holes are more common in the inner disk regions .\\
{\bf 6.}~The outer disks of the two isolated galaxies in our sample, NGC 628 and NGC 6946, appear to be stable with  Q $>$ 1, but the FUV images show ongoing star formation. This suggests that there may be some non-luminous mass or dark matter in their outer disks, which helps increase the disk surface density and supports the formation of local gravitational instabilities. The outer disk star formation can help constrain the total mass in the outer disks of star-forming galaxies. However, in the interacting galaxy NGC 5457, the baryonic surface density is sufficient to trigger local instability.

\acknowledgements
We thank the referee for the valuable suggestions. This publication uses the data from the UVIT, which is part of the AstroSat mission of the Indian Space Research Organisation (ISRO), archived at the Indian Space Science Data Centre (ISSDC). We gratefully thank all the members of various teams for providing support to the project from the early stages of design to launch and observations in the orbit. The authors gratefully acknowledge the IUSSTF grant JC-014/2017, which enabled the authors MD, NNP, and KSD to visit CWRU and develop the science presented in this paper. JY thanks Chayan Mondal and Vikrant Jadhav for productive discussions.\\

\vspace{5mm}

\facilities{Astrosat(UVIT), GALEX, VLA}
\software{Source Extractor\citep{Bertin1996, Barbary2016}, {topcat} \citep{Taylor2005}, 
{IRAF} \citep{Tody1993}, 
Astropy \citep{2013A&A...558A..33A}, 
Matplotlib \citep{Hunter2007}, NumPy \citep{oliphant2006guide}.}

\vspace{5mm}
\bibliography{sample63}{}

\begin{thebibliography}{}
\expandafter\ifx\csname natexlab\endcsname\relax\def\natexlab#1{#1}\fi
\providecommand{\url}[1]{\href{#1}{#1}}
\providecommand{\dodoi}[1]{doi:~\href{http://doi.org/#1}{\nolinkurl{#1}}}
\providecommand{\doeprint}[1]{\href{http://ascl.net/#1}{\nolinkurl{http://ascl.net/#1}}}
\providecommand{\doarXiv}[1]{\href{https://arxiv.org/abs/#1}{\nolinkurl{https://arxiv.org/abs/#1}}}

\bibitem[{{Aniyan} {et~al.}(2018){Aniyan}, {Freeman}, {Arnaboldi}, {Gerhard},
  {Coccato}, {Fabricius}, {Kuijken}, {Merrifield}, \&
  {Ponomareva}}]{Aniyan2018}
{Aniyan}, S., {Freeman}, K.~C., {Arnaboldi}, M., {et~al.} 2018, \mnras, 476,
  1909, \dodoi{10.1093/mnras/sty310}

\bibitem[{{Astropy Collaboration} {et~al.}(2013){Astropy Collaboration},
  {Robitaille}, {Tollerud}, {Greenfield}, {Droettboom}, {Bray}, {Aldcroft},
  {Davis}, {Ginsburg}, {Price-Whelan}, {Kerzendorf}, {Conley}, {Crighton},
  {Barbary}, {Muna}, {Ferguson}, {Grollier}, {Parikh}, {Nair}, {Unther},
  {Deil}, {Woillez}, {Conseil}, {Kramer}, {Turner}, {Singer}, {Fox}, {Weaver},
  {Zabalza}, {Edwards}, {Azalee Bostroem}, {Burke}, {Casey}, {Crawford},
  {Dencheva}, {Ely}, {Jenness}, {Labrie}, {Lim}, {Pierfederici}, {Pontzen},
  {Ptak}, {Refsdal}, {Servillat}, \& {Streicher}}]{2013A&A...558A..33A}
{Astropy Collaboration}, {Robitaille}, T.~P., {Tollerud}, E.~J., {et~al.} 2013,
  \aap, 558, A33, \dodoi{10.1051/0004-6361/201322068}

\bibitem[{{Bagetakos} {et~al.}(2011){Bagetakos}, {Brinks}, {Walter}, {de Blok},
  {Usero}, {Leroy}, {Rich}, \& {Kennicutt}}]{2011AJBagetakos}
{Bagetakos}, I., {Brinks}, E., {Walter}, F., {et~al.} 2011, \aj, 141, 23,
  \dodoi{10.1088/0004-6256/141/1/23}

\bibitem[{Barbary(2016)}]{Barbary2016}
Barbary, K. 2016, Journal of Open Source Software, 1, 58,
  \dodoi{10.21105/joss.00058}

\bibitem[{{Barnes} {et~al.}(2012){Barnes}, {van Zee}, {C{\^o}t{\'e}}, \&
  {Schade}}]{barnes.etal.2012}
{Barnes}, K.~L., {van Zee}, L., {C{\^o}t{\'e}}, S., \& {Schade}, D. 2012, \apj,
  757, 64, \dodoi{10.1088/0004-637X/757/1/64}

\bibitem[{{Barnes} {et~al.}(2011){Barnes}, {van Zee}, \&
  {Skillman}}]{Barnes2011}
{Barnes}, K.~L., {van Zee}, L., \& {Skillman}, E.~D. 2011, \apj, 743, 137,
  \dodoi{10.1088/0004-637X/743/2/137}

\bibitem[{{Belley} \& {Roy}(1992)}]{Belley1992}
{Belley}, J., \& {Roy}, J.-R. 1992, \apjs, 78, 61, \dodoi{10.1086/191621}

\bibitem[{{Bertin} \& {Arnouts}(1996)}]{Bertin1996}
{Bertin}, E., \& {Arnouts}, S. 1996, \aaps, 117, 393,
  \dodoi{10.1051/aas:1996164}

\bibitem[{{Bicalho} {et~al.}(2019){Bicalho}, {Combes}, {Rubio}, {Verdugo}, \&
  {Salome}}]{bicalho.etal.2019}
{Bicalho}, I.~C., {Combes}, F., {Rubio}, M., {Verdugo}, C., \& {Salome}, P.
  2019, \aap, 623, A66, \dodoi{10.1051/0004-6361/201732352}

\bibitem[{{Bigiel} {et~al.}(2010{\natexlab{a}}){Bigiel}, {Leroy}, {Seibert},
  {Walter}, {Blitz}, {Thilker}, \& {Madore}}]{Bigiel2010apjl}
{Bigiel}, F., {Leroy}, A., {Seibert}, M., {et~al.} 2010{\natexlab{a}}, \apjl,
  720, L31, \dodoi{10.1088/2041-8205/720/1/L31}

\bibitem[{{Bigiel} {et~al.}(2010{\natexlab{b}}){Bigiel}, {Leroy}, {Walter},
  {Blitz}, {Brinks}, {de Blok}, \& {Madore}}]{Bigiel2010}
{Bigiel}, F., {Leroy}, A., {Walter}, F., {et~al.} 2010{\natexlab{b}}, \aj, 140,
  1194, \dodoi{10.1088/0004-6256/140/5/1194}

\bibitem[{{Bigiel} {et~al.}(2008){Bigiel}, {Leroy}, {Walter}, {Brinks}, {de
  Blok}, {Madore}, \& {Thornley}}]{bigiel.etal.2008}
---. 2008, \aj, 136, 2846, \dodoi{10.1088/0004-6256/136/6/2846}

\bibitem[{{Binney} \& {Tremaine}(2008)}]{binney.tremaine.2008}
{Binney}, J., \& {Tremaine}, S. 2008, {Galactic Dynamics: Second Edition}

\bibitem[{{Bisnovatyi-Kogan} \& {Silich}(1995)}]{Bisnovatyi1995}
{Bisnovatyi-Kogan}, G.~S., \& {Silich}, S.~A. 1995, Reviews of Modern Physics,
  67, 661, \dodoi{10.1103/RevModPhys.67.661}

\bibitem[{{Boissier} {et~al.}(2008){Boissier}, {Gil de Paz}, {Boselli}, {Buat},
  {Madore}, {Chemin}, {Balkowski}, {Amram}, {Carignan}, \& {van
  Driel}}]{boissier.etal.2008}
{Boissier}, S., {Gil de Paz}, A., {Boselli}, A., {et~al.} 2008, \apj, 681, 244,
  \dodoi{10.1086/588580}

\bibitem[{{Bonaca} {et~al.}(2019){Bonaca}, {Hogg}, {Price-Whelan}, \&
  {Conroy}}]{bonaca.etal.2019}
{Bonaca}, A., {Hogg}, D.~W., {Price-Whelan}, A.~M., \& {Conroy}, C. 2019, \apj,
  880, 38, \dodoi{10.3847/1538-4357/ab2873}

\bibitem[{{Bonnell} {et~al.}(2013){Bonnell}, {Dobbs}, \&
  {Smith}}]{bonnell.etal.2013}
{Bonnell}, I.~A., {Dobbs}, C.~L., \& {Smith}, R.~J. 2013, \mnras, 430, 1790,
  \dodoi{10.1093/mnras/stt004}

\bibitem[{{Boomsma} {et~al.}(2008){Boomsma}, {Oosterloo}, {Fraternali}, {van
  der Hulst}, \& {Sancisi}}]{Boomsma2008}
{Boomsma}, R., {Oosterloo}, T.~A., {Fraternali}, F., {van der Hulst}, J.~M., \&
  {Sancisi}, R. 2008, \aap, 490, 555, \dodoi{10.1051/0004-6361:200810120}

\bibitem[{{Bosma}(1981)}]{bosma.1981}
{Bosma}, A. 1981, \aj, 86, 1791, \dodoi{10.1086/113062}

\bibitem[{{Braun}(1997)}]{braun.1997}
{Braun}, R. 1997, \apj, 484, 637, \dodoi{10.1086/304346}

\bibitem[{{Catinella} {et~al.}(2010){Catinella}, {Schiminovich}, {Kauffmann},
  {Fabello}, {Wang}, {Hummels}, {Lemonias}, {Moran}, {Wu}, {Giovanelli},
  {Haynes}, {Heckman}, {Basu-Zych}, {Blanton}, {Brinchmann}, {Budav{\'a}ri},
  {Gon{\c{c}}alves}, {Johnson}, {Kennicutt}, {Madore}, {Martin}, {Rich},
  {Tacconi}, {Thilker}, {Wild}, \& {Wyder}}]{Catinella2010}
{Catinella}, B., {Schiminovich}, D., {Kauffmann}, G., {et~al.} 2010, \mnras,
  403, 683, \dodoi{10.1111/j.1365-2966.2009.16180.x}

\bibitem[{{Cedr{\'e}s} {et~al.}(2012){Cedr{\'e}s}, {Cepa}, {Bongiovanni},
  {Casta{\~n}eda}, {S{\'a}nchez-Portal}, \& {Tomita}}]{Cedres2012}
{Cedr{\'e}s}, B., {Cepa}, J., {Bongiovanni}, {\'A}., {et~al.} 2012, \aap, 545,
  A43, \dodoi{10.1051/0004-6361/201219571}

\bibitem[{{Cedr{\'e}s} {et~al.}(2013){Cedr{\'e}s}, {Cepa}, {Bongiovanni},
  {Casta{\~n}eda}, {S{\'a}nchez-Portal}, \& {Tomita}}]{Bernabe2013}
---. 2013, \aap, 560, A59, \dodoi{10.1051/0004-6361/201321588}

\bibitem[{{Condon}(1987)}]{Condon1987}
{Condon}, J.~J. 1987, \apjs, 65, 485, \dodoi{10.1086/191234}

\bibitem[{{Cornett} {et~al.}(1994){Cornett}, {O'Connell}, {Greason},
  {Offenberg}, {Angione}, {Bohlin}, {Cheng}, {Roberts}, {Smith}, {Smith},
  {Talbert}, \& {Stecher}}]{Cornett1994}
{Cornett}, R.~H., {O'Connell}, R.~W., {Greason}, M.~R., {et~al.} 1994, \apj,
  426, 553, \dodoi{10.1086/174092}

\bibitem[{{Das} {et~al.}(2010){Das}, {Boone}, \& {Viallefond}}]{das.etal.2010}
{Das}, M., {Boone}, F., \& {Viallefond}, F. 2010, \aap, 523, A63,
  \dodoi{10.1051/0004-6361/200913794}

\bibitem[{{Das} {et~al.}(2020){Das}, {McGaugh}, {Ianjamasimanana}, {Schombert},
  \& {Dwarakanath}}]{das.etal.2020}
{Das}, M., {McGaugh}, S.~S., {Ianjamasimanana}, R., {Schombert}, J., \&
  {Dwarakanath}, K.~S. 2020, \apj, 889, 10, \dodoi{10.3847/1538-4357/ab5fcd}

\bibitem[{{Das} {et~al.}(2006){Das}, {O'Neil}, {Vogel}, \&
  {McGaugh}}]{das.etal.2006}
{Das}, M., {O'Neil}, K., {Vogel}, S.~N., \& {McGaugh}, S. 2006, \apj, 651, 853,
  \dodoi{10.1086/507410}

\bibitem[{{Das} {et~al.}(2019){Das}, {Sengupta}, \& {Honey}}]{das.etal.2019}
{Das}, M., {Sengupta}, C., \& {Honey}, M. 2019, \apj, 871, 197,
  \dodoi{10.3847/1538-4357/aaf864}

\bibitem[{{Dawson} {et~al.}(2013){Dawson}, {McClure-Griffiths}, {Fukui},
  {Dickey}, {Wong}, {Hughes}, \& {Kawamura}}]{Dawson2013}
{Dawson}, J.~R., {McClure-Griffiths}, N.~M., {Fukui}, Y., {et~al.} 2013, in IAU
  Symposium, Vol. 292, Molecular Gas, Dust, and Star Formation in Galaxies, ed.
  T.~{Wong} \& J.~{Ott}, 83--86, \dodoi{10.1017/S1743921313000525}

\bibitem[{{Dawson} {et~al.}(2011){Dawson}, {McClure-Griffiths}, {Kawamura},
  {Mizuno}, {Onishi}, {Mizuno}, \& {Fukui}}]{Dawson2011}
{Dawson}, J.~R., {McClure-Griffiths}, N.~M., {Kawamura}, A., {et~al.} 2011,
  \apj, 728, 127, \dodoi{10.1088/0004-637X/728/2/127}

\bibitem[{{de Blok} {et~al.}(2008){de Blok}, {Walter}, {Brinks},
  {Trachternach}, {Oh}, \& {Kennicutt}}]{deblok08}
{de Blok}, W.~J.~G., {Walter}, F., {Brinks}, E., {et~al.} 2008, \aj, 136, 2648,
  \dodoi{10.1088/0004-6256/136/6/2648}

\bibitem[{{Dekel} \& {Birnboim}(2006)}]{Dekel2006}
{Dekel}, A., \& {Birnboim}, Y. 2006, \mnras, 368, 2,
  \dodoi{10.1111/j.1365-2966.2006.10145.x}

\bibitem[{{Dekel} {et~al.}(2009){Dekel}, {Birnboim}, {Engel}, {Freundlich},
  {Goerdt}, {Mumcuoglu}, {Neistein}, {Pichon}, {Teyssier}, \&
  {Zinger}}]{Dekel2009}
{Dekel}, A., {Birnboim}, Y., {Engel}, G., {et~al.} 2009, \nat, 457, 451,
  \dodoi{10.1038/nature07648}

\bibitem[{{Dessauges-Zavadsky} {et~al.}(2014){Dessauges-Zavadsky}, {Verdugo},
  {Combes}, \& {Pfenniger}}]{dessauges-zavadsky.2014}
{Dessauges-Zavadsky}, M., {Verdugo}, C., {Combes}, F., \& {Pfenniger}, D. 2014,
  \aap, 566, A147, \dodoi{10.1051/0004-6361/201323330}

\bibitem[{{Donas} \& {Deharveng}(1984)}]{Donas1984}
{Donas}, J., \& {Deharveng}, J.~M. 1984, \aap, 140, 325

\bibitem[{{Doyle} \& {Drinkwater}(2006)}]{Doyle2006}
{Doyle}, M.~T., \& {Drinkwater}, M.~J. 2006, \mnras, 372, 977,
  \dodoi{10.1111/j.1365-2966.2006.10931.x}

\bibitem[{{Ehlerov{\'a}} \& {Palou{\v{s}}}(2002)}]{Ehlerov2002}
{Ehlerov{\'a}}, S., \& {Palou{\v{s}}}, J. 2002, \mnras, 330, 1022,
  \dodoi{10.1046/j.1365-8711.2002.05156.x}

\bibitem[{{Eldridge} \& {Xiao}(2019)}]{Eldridge2019}
{Eldridge}, J.~J., \& {Xiao}, L. 2019, \mnras, 485, L58,
  \dodoi{10.1093/mnrasl/slz030}

\bibitem[{{Elmegreen}(1998)}]{Elmegreen1998}
{Elmegreen}, D.~M. 1998, {Galaxies and galactic structure}

\bibitem[{{Elmegreen} \& {Elmegreen}(1984)}]{Elmegreen1984}
{Elmegreen}, D.~M., \& {Elmegreen}, B.~G. 1984, \apjs, 54, 127,
  \dodoi{10.1086/190922}

\bibitem[{{Fan} {et~al.}(2013){Fan}, {Katz}, {Randall}, \&
  {Reece}}]{fan.etal.2013}
{Fan}, J., {Katz}, A., {Randall}, L., \& {Reece}, M. 2013, Physics of the Dark
  Universe, 2, 139, \dodoi{10.1016/j.dark.2013.07.001}

\bibitem[{{Ferguson} {et~al.}(1998{\natexlab{a}}){Ferguson}, {Gallagher}, \&
  {Wyse}}]{ferguson.etal.1998AJ}
{Ferguson}, A. M.~N., {Gallagher}, J.~S., \& {Wyse}, R. F.~G.
  1998{\natexlab{a}}, \aj, 116, 673, \dodoi{10.1086/300456}

\bibitem[{{Ferguson} {et~al.}(1998{\natexlab{b}}){Ferguson}, {Wyse},
  {Gallagher}, \& {Hunter}}]{Ferguson1998}
{Ferguson}, A. M.~N., {Wyse}, R. F.~G., {Gallagher}, J.~S., \& {Hunter}, D.~A.
  1998{\natexlab{b}}, \apjl, 506, L19, \dodoi{10.1086/311626}

\bibitem[{{Fitzpatrick}(1999)}]{fitzpatrick1999}
{Fitzpatrick}, E.~L. 1999, \pasp, 111, 63, \dodoi{10.1086/316293}

\bibitem[{{Gil de Paz} {et~al.}(2005){Gil de Paz}, {Madore}, {Boissier},
  {Swaters}, {Popescu}, {Tuffs}, {Sheth}, {Kennicutt}, {Bianchi}, {Thilker}, \&
  {Martin}}]{Gil2005}
{Gil de Paz}, A., {Madore}, B.~F., {Boissier}, S., {et~al.} 2005, \apjl, 627,
  L29, \dodoi{10.1086/432054}

\bibitem[{{Gil de Paz} {et~al.}(2007){Gil de Paz}, {Madore}, {Boissier},
  {Thilker}, {Bianchi}, {S{\'a}nchez Contreras}, {Barlow}, {Conrow}, {Forster},
  {Friedman}, {Martin}, {Morrissey}, {Neff}, {Rich}, {Schiminovich}, {Seibert},
  {Small}, {Donas}, {Heckman}, {Lee}, {Milliard}, {Szalay}, {Wyder}, \&
  {Yi}}]{Gil2007}
---. 2007, \apj, 661, 115, \dodoi{10.1086/513730}

\bibitem[{{Glover} \& {Clark}(2012)}]{glover.clark.2012}
{Glover}, S. C.~O., \& {Clark}, P.~C. 2012, \mnras, 421, 9,
  \dodoi{10.1111/j.1365-2966.2011.19648.x}

\bibitem[{{Goddard} {et~al.}(2011){Goddard}, {Bresolin}, {Kennicutt},
  {Ryan-Weber}, \& {Rosales-Ortega}}]{goddard.etal.2011}
{Goddard}, Q.~E., {Bresolin}, F., {Kennicutt}, R.~C., {Ryan-Weber}, E.~V., \&
  {Rosales-Ortega}, F.~F. 2011, \mnras, 412, 1246,
  \dodoi{10.1111/j.1365-2966.2010.17990.x}

\bibitem[{{Gu{\'e}lin} \& {Weliachew}(1970)}]{Gulin1970}
{Gu{\'e}lin}, M., \& {Weliachew}, L. 1970, \aap, 7, 141

\bibitem[{{Hitschfeld} {et~al.}(2009){Hitschfeld}, {Kramer}, {Schuster},
  {Garcia-Burillo}, \& {Stutzki}}]{Hitschfeld2009}
{Hitschfeld}, M., {Kramer}, C., {Schuster}, K.~F., {Garcia-Burillo}, S., \&
  {Stutzki}, J. 2009, \aap, 495, 795, \dodoi{10.1051/0004-6361:200810898}

\bibitem[{{Ho} {et~al.}(2019){Ho}, {Martin}, \& {Turner}}]{ho.etal.2019}
{Ho}, S.~H., {Martin}, C.~L., \& {Turner}, M.~L. 2019, \apj, 875, 54,
  \dodoi{10.3847/1538-4357/ab0ec2}

\bibitem[{{Huang} {et~al.}(2012){Huang}, {Haynes}, {Giovanelli}, \&
  {Brinchmann}}]{Huang2012}
{Huang}, S., {Haynes}, M.~P., {Giovanelli}, R., \& {Brinchmann}, J. 2012, \apj,
  756, 113, \dodoi{10.1088/0004-637X/756/2/113}

\bibitem[{{Hunter} {et~al.}(2016){Hunter}, {Elmegreen}, \&
  {Gehret}}]{2016Hunter}
{Hunter}, D.~A., {Elmegreen}, B.~G., \& {Gehret}, E. 2016, \aj, 151, 136,
  \dodoi{10.3847/0004-6256/151/6/136}

\bibitem[{{Hunter}(2007)}]{Hunter2007}
{Hunter}, J.~D. 2007, Computing in Science and Engineering, 9, 90,
  \dodoi{10.1109/MCSE.2007.55}

\bibitem[{{Hwang} {et~al.}(2019){Hwang}, {Barrera-Ballesteros}, {Heckman},
  {Rowlands}, {Lin}, {Rodriguez-Gomez}, {Pan}, {Hsieh}, {S{\'a}nchez},
  {Bizyaev}, {S{\'a}nchez Almeida}, {Thilker}, {Lotz}, {Jones}, {Nair},
  {Andrews}, \& {Drory}}]{Hwang2019}
{Hwang}, H.-C., {Barrera-Ballesteros}, J.~K., {Heckman}, T.~M., {et~al.} 2019,
  \apj, 872, 144, \dodoi{10.3847/1538-4357/aaf7a3}

\bibitem[{{Hyman} {et~al.}(2000){Hyman}, {Lacey}, {Weiler}, \& {Van
  Dyk}}]{Hyman2000}
{Hyman}, S.~D., {Lacey}, C.~K., {Weiler}, K.~W., \& {Van Dyk}, S.~D. 2000, \aj,
  119, 1711, \dodoi{10.1086/301306}

\bibitem[{{Ianjamasimanana} {et~al.}(2012){Ianjamasimanana}, {de Blok},
  {Walter}, \& {Heald}}]{ianjamasimanana12}
{Ianjamasimanana}, R., {de Blok}, W.~J.~G., {Walter}, F., \& {Heald}, G.~H.
  2012, \aj, 144, 96, \dodoi{10.1088/0004-6256/144/4/96}

\bibitem[{{Kamphuis} \& {Briggs}(1992)}]{Kamphuis1992}
{Kamphuis}, J., \& {Briggs}, F. 1992, \aap, 253, 335

\bibitem[{{Kamphuis}(1993)}]{Kamphuis1993}
{Kamphuis}, J.~J. 1993, PhD thesis, -

\bibitem[{{Kennicutt}(1998{\natexlab{a}})}]{Kennicutt1998}
{Kennicutt}, Robert~C., J. 1998{\natexlab{a}}, \apj, 498, 541,
  \dodoi{10.1086/305588}

\bibitem[{{Kennicutt}(1998{\natexlab{b}})}]{kenni1998}
---. 1998{\natexlab{b}}, \araa, 36, 189, \dodoi{10.1146/annurev.astro.36.1.189}

\bibitem[{{Kennicutt} \& {Evans}(2012)}]{kennicutt.evans.2012}
{Kennicutt}, R.~C., \& {Evans}, N.~J. 2012, \araa, 50, 531,
  \dodoi{10.1146/annurev-astro-081811-125610}

\bibitem[{{Kim} {et~al.}(1999){Kim}, {Dopita}, {Staveley-Smith}, \&
  {Bessell}}]{Kim1999}
{Kim}, S., {Dopita}, M.~A., {Staveley-Smith}, L., \& {Bessell}, M.~S. 1999,
  \aj, 118, 2797, \dodoi{10.1086/301116}

\bibitem[{{Knapen} {et~al.}(1996){Knapen}, {Beckman}, {Cepa}, \&
  {Nakai}}]{Knapen1996}
{Knapen}, J.~H., {Beckman}, J.~E., {Cepa}, J., \& {Nakai}, N. 1996, \aap, 308,
  27.
\newblock \doarXiv{astro-ph/9509076}

\bibitem[{{Koopmann} \& {Kenney}(2004)}]{koopman.kenney.2004}
{Koopmann}, R.~A., \& {Kenney}, J. D.~P. 2004, \apj, 613, 866,
  \dodoi{10.1086/423191}

\bibitem[{{Kormendy}(2013)}]{Kormendy2013}
{Kormendy}, J. 2013, {Secular Evolution in Disk Galaxies}, ed.
  J.~{Falc{\'o}n-Barroso} \& J.~H. {Knapen}, 1

\bibitem[{{Koyama} {et~al.}(1986){Koyama}, {Ikeuchi}, \&
  {Tomisaka}}]{Tomisaka1986}
{Koyama}, K., {Ikeuchi}, S., \& {Tomisaka}, K. 1986, \pasj, 38, 503

\bibitem[{{Krumholz} \& {McKee}(2008)}]{krumholz.mckee.2008}
{Krumholz}, M.~R., \& {McKee}, C.~F. 2008, \nat, 451, 1082,
  \dodoi{10.1038/nature06620}

\bibitem[{{Kumar} {et~al.}(2012){Kumar}, {Ghosh}, {Hutchings}, {Kamath},
  {Kathiravan}, {Mahesh}, {Murthy}, {Nagbhushana}, {Pati}, {Rao}, {Rao},
  {Sriram}, \& {Tandon}}]{Kumar2012}
{Kumar}, A., {Ghosh}, S.~K., {Hutchings}, J., {et~al.} 2012, Society of
  Photo-Optical Instrumentation Engineers (SPIE) Conference Series, Vol. 8443,
  {Ultra Violet Imaging Telescope (UVIT) on ASTROSAT}, 84431N,
  \dodoi{10.1117/12.924507}

\bibitem[{{Leitherer}(1998)}]{Leitherer1998}
{Leitherer}, C. 1998, Astronomical Society of the Pacific Conference Series,
  Vol. 142, {The Initial Mass Function in Starburst Galaxies}, ed. G.~{Gilmore}
  \& D.~{Howell}, 61

\bibitem[{{Leitherer} {et~al.}(1999){Leitherer}, {Schaerer}, {Goldader},
  {Delgado}, {Robert}, {Kune}, {de Mello}, {Devost}, \&
  {Heckman}}]{Leitherer1999}
{Leitherer}, C., {Schaerer}, D., {Goldader}, J.~D., {et~al.} 1999, \apjs, 123,
  3, \dodoi{10.1086/313233}

\bibitem[{{Lemonias} {et~al.}(2011){Lemonias}, {Schiminovich}, {Thilker},
  {Wyder}, {Martin}, {Seibert}, {Treyer}, {Bianchi}, {Heckman}, {Madore}, \&
  {Rich}}]{lemonias.etal.2011}
{Lemonias}, J.~J., {Schiminovich}, D., {Thilker}, D., {et~al.} 2011, \apj, 733,
  74, \dodoi{10.1088/0004-637X/733/2/74}

\bibitem[{{Leroy} {et~al.}(2008){Leroy}, {Walter}, {Brinks}, {Bigiel}, {de
  Blok}, {Madore}, \& {Thornley}}]{Leroy2008}
{Leroy}, A.~K., {Walter}, F., {Brinks}, E., {et~al.} 2008, \aj, 136, 2782,
  \dodoi{10.1088/0004-6256/136/6/2782}

\bibitem[{{Lira} {et~al.}(2007){Lira}, {Johnson}, {Lawrence}, \& {Cid
  Fernandes}}]{Lira2007}
{Lira}, P., {Johnson}, R.~A., {Lawrence}, A., \& {Cid Fernandes}, R. 2007,
  \mnras, 382, 1552, \dodoi{10.1111/j.1365-2966.2007.12006.x}

\bibitem[{{Mac Low} \& {McCray}(1988)}]{MacLow1988}
{Mac Low}, M.-M., \& {McCray}, R. 1988, \apj, 324, 776, \dodoi{10.1086/165936}

\bibitem[{{MacLow} {et~al.}(1986){MacLow}, {McCray}, \&
  {Kafatos}}]{Richard1986}
{MacLow}, M.~M., {McCray}, R., \& {Kafatos}, M. 1986, \pasp, 98, 1104,
  \dodoi{10.1086/131895}

\bibitem[{{Maddox} {et~al.}(2015){Maddox}, {Hess}, {Obreschkow}, {Jarvis}, \&
  {Blyth}}]{Maddox2015}
{Maddox}, N., {Hess}, K.~M., {Obreschkow}, D., {Jarvis}, M.~J., \& {Blyth},
  S.~L. 2015, \mnras, 447, 1610, \dodoi{10.1093/mnras/stu2532}

\bibitem[{{Martinet} \& {Friedli}(1997)}]{martinet.friedli.1997}
{Martinet}, L., \& {Friedli}, D. 1997, \aap, 323, 363.
\newblock \doarXiv{astro-ph/9701091}

\bibitem[{{McCray} \& {Kafatos}(1987)}]{McCray1987}
{McCray}, R., \& {Kafatos}, M. 1987, \apj, 317, 190, \dodoi{10.1086/165267}

\bibitem[{{McGaugh} {et~al.}(2017){McGaugh}, {Schombert}, \&
  {Lelli}}]{mcgaugh.etal.2017}
{McGaugh}, S.~S., {Schombert}, J.~M., \& {Lelli}, F. 2017, \apj, 851, 22,
  \dodoi{10.3847/1538-4357/aa9790}

\bibitem[{{Mihos} {et~al.}(2013){Mihos}, {Harding}, {Spengler}, {Rudick}, \&
  {Feldmeier}}]{Mihos2013}
{Mihos}, J.~C., {Harding}, P., {Spengler}, C.~E., {Rudick}, C.~S., \&
  {Feldmeier}, J.~J. 2013, \apj, 762, 82, \dodoi{10.1088/0004-637X/762/2/82}

\bibitem[{{Mondal} {et~al.}(2019){Mondal}, {Subramaniam}, \&
  {George}}]{Chayan2019}
{Mondal}, C., {Subramaniam}, A., \& {George}, K. 2019, \aj, 158, 229,
  \dodoi{10.3847/1538-3881/ab4ea1}

\bibitem[{{Morrissey} {et~al.}(2007){Morrissey}, {Conrow}, {Barlow}, {Small},
  {Seibert}, {Wyder}, {Budav{\'a}ri}, {Arnouts}, {Friedman}, {Forster},
  {Martin}, {Neff}, {Schiminovich}, {Bianchi}, {Donas}, {Heckman}, {Lee},
  {Madore}, {Milliard}, {Rich}, {Szalay}, {Welsh}, \& {Yi}}]{Morrissey2007}
{Morrissey}, P., {Conrow}, T., {Barlow}, T.~A., {et~al.} 2007, \apjs, 173, 682,
  \dodoi{10.1086/520512}

\bibitem[{{Mulcahy} {et~al.}(2017){Mulcahy}, {Beck}, \& {Heald}}]{Mulcahy2017}
{Mulcahy}, D.~D., {Beck}, R., \& {Heald}, G.~H. 2017, \aap, 600, A6,
  \dodoi{10.1051/0004-6361/201629907}

\bibitem[{{Murthy} {et~al.}(2016){Murthy}, {Rahna}, {Safonova}, {Sutaria},
  {Gudennavar}, \& {Bubbly}}]{Jayanth2016}
{Murthy}, J., {Rahna}, P.~T., {Safonova}, M., {et~al.} 2016, {JUDE: An
  Utraviolet Imaging Telescope pipeline}.
\newblock \doeprint{1607.007}

\bibitem[{{Murthy} {et~al.}(2017){Murthy}, {Rahna}, {Sutaria}, {Safonova},
  {Gudennavar}, \& {Bubbly}}]{jayanth2017}
{Murthy}, J., {Rahna}, P.~T., {Sutaria}, F., {et~al.} 2017, Astronomy and
  Computing, 20, 120, \dodoi{10.1016/j.ascom.2017.07.001}

\bibitem[{{Obreschkow} {et~al.}(2016){Obreschkow}, {Glazebrook}, {Kilborn}, \&
  {Lutz}}]{Obreschkow2016}
{Obreschkow}, D., {Glazebrook}, K., {Kilborn}, V., \& {Lutz}, K. 2016, \apjl,
  824, L26, \dodoi{10.3847/2041-8205/824/2/L26}

\bibitem[{Oliphant(2015)}]{oliphant2006guide}
Oliphant, T.~E. 2015, Guide to NumPy, 2nd edn. (North Charleston, SC, USA:
  CreateSpace Independent Publishing Platform)

\bibitem[{{Olling}(1996)}]{olling.1996}
{Olling}, R.~P. 1996, \aj, 112, 481, \dodoi{10.1086/118029}

\bibitem[{{Patra}(2018)}]{patra18c}
{Patra}, N.~N. 2018, \mnras, 480, 4369, \dodoi{10.1093/mnras/sty2167}

\bibitem[{{Patra}(2020{\natexlab{a}})}]{patra20a}
---. 2020{\natexlab{a}}, \mnras, 495, 2867, \dodoi{10.1093/mnras/staa1353}

\bibitem[{{Patra}(2020{\natexlab{b}})}]{patra20b}
---. 2020{\natexlab{b}}, \aap, 638, A66, \dodoi{10.1051/0004-6361/201936483}

\bibitem[{{Patra} {et~al.}(2016){Patra}, {Chengalur}, {Karachentsev}, {Kaisin},
  \& {Begum}}]{patra.etal.2016}
{Patra}, N.~N., {Chengalur}, J.~N., {Karachentsev}, I.~D., {Kaisin}, S.~S., \&
  {Begum}, A. 2016, \mnras, 456, 2467, \dodoi{10.1093/mnras/stv2789}

\bibitem[{{Pearson} {et~al.}(2019){Pearson}, {Wang}, {Alpaslan}, {Baldry},
  {Bilicki}, {Brown}, {Grootes}, {Holwerda}, {Kitching}, {Kruk}, \& {van der
  Tak}}]{Pearson2019}
{Pearson}, W.~J., {Wang}, L., {Alpaslan}, M., {et~al.} 2019, \aap, 631, A51,
  \dodoi{10.1051/0004-6361/201936337}

\bibitem[{{Postma} \& {Leahy}(2017)}]{Joe2017}
{Postma}, J.~E., \& {Leahy}, D. 2017, \pasp, 129, 115002,
  \dodoi{10.1088/1538-3873/aa8800}

\bibitem[{{Prochaska} \& {Wolfe}(2009)}]{Prochaska2009}
{Prochaska}, J.~X., \& {Wolfe}, A.~M. 2009, \apj, 696, 1543,
  \dodoi{10.1088/0004-637X/696/2/1543}

\bibitem[{{Rahna} {et~al.}(2018){Rahna}, {Das}, {Murthy}, {Gudennavar}, \&
  {Bubbly}}]{rahna.etal.2018}
{Rahna}, P.~T., {Das}, M., {Murthy}, J., {Gudennavar}, S.~B., \& {Bubbly},
  S.~G. 2018, \mnras, 481, 1212, \dodoi{10.1093/mnras/sty2250}

\bibitem[{{Salim} {et~al.}(2007){Salim}, {Rich}, {Charlot}, {Brinchmann},
  {Johnson}, {Schiminovich}, {Seibert}, {Mallery}, {Heckman}, {Forster},
  {Friedman}, {Martin}, {Morrissey}, {Neff}, {Small}, {Wyder}, {Bianchi},
  {Donas}, {Lee}, {Madore}, {Milliard}, {Szalay}, {Welsh}, \& {Yi}}]{Salim2007}
{Salim}, S., {Rich}, R.~M., {Charlot}, S., {et~al.} 2007, \apjs, 173, 267,
  \dodoi{10.1086/519218}

\bibitem[{{Sandage} \& {Bedke}(1994)}]{Sandage1994}
{Sandage}, A., \& {Bedke}, J. 1994, {The Carnegie atlas of galaxies}, Vol. 638

\bibitem[{{Saponara} {et~al.}(2020){Saponara}, {Koribalski}, {Patra}, \&
  {Benaglia}}]{saponara20}
{Saponara}, J., {Koribalski}, B.~S., {Patra}, N.~N., \& {Benaglia}, P. 2020,
  \apss, 365, 111, \dodoi{10.1007/s10509-020-03825-2}

\bibitem[{{Scalo}(1998)}]{Scalo1998}
{Scalo}, J. 1998, Astronomical Society of the Pacific Conference Series, Vol.
  142, {The IMF Revisited: A Case for Variations}, ed. G.~{Gilmore} \&
  D.~{Howell}, 201

\bibitem[{{Schlafly} \& {Finkbeiner}(2011)}]{Schlafly2011}
{Schlafly}, E.~F., \& {Finkbeiner}, D.~P. 2011, \apj, 737, 103,
  \dodoi{10.1088/0004-637X/737/2/103}

\bibitem[{{Schmitt} {et~al.}(2006){Schmitt}, {Calzetti}, {Armus}, {Giavalisco},
  {Heckman}, {Kennicutt}, {Leitherer}, \& {Meurer}}]{Schmitt2006}
{Schmitt}, H.~R., {Calzetti}, D., {Armus}, L., {et~al.} 2006, \apj, 643, 173,
  \dodoi{10.1086/501512}

\bibitem[{{Schombert} {et~al.}(2011){Schombert}, {Maciel}, \&
  {McGaugh}}]{schombert.etal.2011}
{Schombert}, J., {Maciel}, T., \& {McGaugh}, S. 2011, Advances in Astronomy,
  2011, 143698, \dodoi{10.1155/2011/143698}

\bibitem[{{Schruba} {et~al.}(2011){Schruba}, {Leroy}, {Walter}, {Bigiel},
  {Brinks}, {de Blok}, {Dumas}, {Kramer}, {Rosolowsky}, {Sandstrom},
  {Schuster}, {Usero}, {Weiss}, \& {Wiesemeyer}}]{schruba11}
{Schruba}, A., {Leroy}, A.~K., {Walter}, F., {et~al.} 2011, \aj, 142, 37,
  \dodoi{10.1088/0004-6256/142/2/37}

\bibitem[{{Sharina} {et~al.}(1997){Sharina}, {Karachentsev}, \&
  {Tikhonov}}]{Sharina1997}
{Sharina}, M.~E., {Karachentsev}, I.~D., \& {Tikhonov}, N.~A. 1997, Astronomy
  Letters, 23, 373

\bibitem[{{Shi} {et~al.}(2018){Shi}, {Yan}, {Armus}, {Gu}, {Helou}, {Qiu},
  {Gwyn}, {Stierwalt}, {Fang}, {Chen}, {Zhou}, {Wu}, {Zheng}, {Zhang}, {Gao},
  \& {Wang}}]{shi.etal.2018}
{Shi}, Y., {Yan}, L., {Armus}, L., {et~al.} 2018, \apj, 853, 149,
  \dodoi{10.3847/1538-4357/aaa3e6}

\bibitem[{{Silich} {et~al.}(1996){Silich}, {Franco}, {Palous}, \&
  {Tenorio-Tagle}}]{Silich1996}
{Silich}, S.~A., {Franco}, J., {Palous}, J., \& {Tenorio-Tagle}, G. 1996, \apj,
  468, 722, \dodoi{10.1086/177728}

\bibitem[{{Silk} \& {Mamon}(2012)}]{Silk2012}
{Silk}, J., \& {Mamon}, G.~A. 2012, Research in Astronomy and Astrophysics, 12,
  917, \dodoi{10.1088/1674-4527/12/8/004}

\bibitem[{{Sofue} {et~al.}(1999){Sofue}, {Tutui}, {Honma}, {Tomita},
  {Takamiya}, {Koda}, \& {Takeda}}]{Sofue1999}
{Sofue}, Y., {Tutui}, Y., {Honma}, M., {et~al.} 1999, \apj, 523, 136,
  \dodoi{10.1086/307731}

\bibitem[{{Stanimirovic} {et~al.}(1999){Stanimirovic}, {Staveley-Smith},
  {Dickey}, {Sault}, \& {Snowden}}]{Stanimirovic1999}
{Stanimirovic}, S., {Staveley-Smith}, L., {Dickey}, J.~M., {Sault}, R.~J., \&
  {Snowden}, S.~L. 1999, \mnras, 302, 417,
  \dodoi{10.1046/j.1365-8711.1999.02013.x}

\bibitem[{{Tandon} {et~al.}(2017){Tandon}, {Subramaniam}, {Girish}, {Postma},
  {Sankarasubramanian}, {Sriram}, {Stalin}, {Mondal}, {Sahu}, {Joseph},
  {Hutchings}, {Ghosh}, {Barve}, {George}, {Kamath}, {Kathiravan}, {Kumar},
  {Lancelot}, {Leahy}, {Mahesh}, {Mohan}, {Nagabhushana}, {Pati}, {Kameswara
  Rao}, {Sreedhar}, \& {Sreekumar}}]{Tandon2017}
{Tandon}, S.~N., {Subramaniam}, A., {Girish}, V., {et~al.} 2017, \aj, 154, 128,
  \dodoi{10.3847/1538-3881/aa8451}

\bibitem[{{Taylor}(2005)}]{Taylor2005}
{Taylor}, M.~B. 2005, Astronomical Society of the Pacific Conference Series,
  Vol. 347, {TOPCAT \&amp; STIL: Starlink Table/VOTable Processing Software},
  ed. P.~{Shopbell}, M.~{Britton}, \& R.~{Ebert}, 29

\bibitem[{{Thilker} {et~al.}(2005){Thilker}, {Bianchi}, {Boissier}, {Gil de
  Paz}, {Madore}, {Martin}, {Meurer}, {Neff}, {Rich}, {Schiminovich},
  {Seibert}, {Wyder}, {Barlow}, {Byun}, {Donas}, {Forster}, {Friedman},
  {Heckman}, {Jelinsky}, {Lee}, {Malina}, {Milliard}, {Morrissey}, {Siegmund},
  {Small}, {Szalay}, \& {Welsh}}]{Thilker2005}
{Thilker}, D.~A., {Bianchi}, L., {Boissier}, S., {et~al.} 2005, \apjl, 619,
  L79, \dodoi{10.1086/425251}

\bibitem[{{Thilker} {et~al.}(2007){Thilker}, {Bianchi}, {Meurer}, {Gil de Paz},
  {Boissier}, {Madore}, {Boselli}, {Ferguson}, {Mu{\~n}oz-Mateos}, {Madsen},
  {Hameed}, {Overzier}, {Forster}, {Friedman}, {Martin}, {Morrissey}, {Neff},
  {Schiminovich}, {Seibert}, {Small}, {Wyder}, {Donas}, {Heckman}, {Lee},
  {Milliard}, {Rich}, {Szalay}, {Welsh}, \& {Yi}}]{Thilker2007}
{Thilker}, D.~A., {Bianchi}, L., {Meurer}, G., {et~al.} 2007, \apjs, 173, 538,
  \dodoi{10.1086/523853}

\bibitem[{{Tody}(1993)}]{Tody1993}
{Tody}, D. 1993, Astronomical Society of the Pacific Conference Series,
  Vol.~52, {IRAF in the Nineties}, ed. R.~J. {Hanisch}, R.~J.~V. {Brissenden},
  \& J.~{Barnes}, 173

\bibitem[{{Toomre}(1964)}]{toomre64}
{Toomre}, A. 1964, \apj, 139, 1217, \dodoi{10.1086/147861}

\bibitem[{{Trewhella}(1998)}]{Trewhella1998}
{Trewhella}, M. 1998, \mnras, 297, 807,
  \dodoi{10.1046/j.1365-8711.1998.01523.x}

\bibitem[{{Tully}(1988)}]{Tully1988}
{Tully}, R.~B. 1988, Science, 242, 310

\bibitem[{{van der Hulst} \& {Sancisi}(1988)}]{Hulst1988}
{van der Hulst}, T., \& {Sancisi}, R. 1988, \aj, 95, 1354,
  \dodoi{10.1086/114731}

\bibitem[{{Walter} {et~al.}(2008){Walter}, {Brinks}, {de Blok}, {Bigiel},
  {Kennicutt}, {Thornley}, \& {Leroy}}]{Walter2008}
{Walter}, F., {Brinks}, E., {de Blok}, W.~J.~G., {et~al.} 2008, \aj, 136, 2563,
  \dodoi{10.1088/0004-6256/136/6/2563}

\bibitem[{{Wyder} {et~al.}(2009){Wyder}, {Martin}, {Barlow}, {Foster},
  {Friedman}, {Morrissey}, {Neff}, {Neill}, {Schiminovich}, {Seibert},
  {Bianchi}, {Donas}, {Heckman}, {Lee}, {Madore}, {Milliard}, {Rich}, {Szalay},
  \& {Yi}}]{wyder.etal.2009}
{Wyder}, T.~K., {Martin}, D.~C., {Barlow}, T.~A., {et~al.} 2009, \apj, 696,
  1834, \dodoi{10.1088/0004-637X/696/2/1834}

\bibitem[{{Yates} {et~al.}(2012){Yates}, {Kauffmann}, \& {Guo}}]{Guo2012}
{Yates}, R.~M., {Kauffmann}, G., \& {Guo}, Q. 2012, \mnras, 422, 215,
  \dodoi{10.1111/j.1365-2966.2012.20595.x}

\bibitem[{{Young} {et~al.}(1986){Young}, {Kenney}, {Tacconi}, {Claussen},
  {Huang}, {Tacconi-Garman}, {Xie}, \& {Schloerb}}]{Young1986}
{Young}, J.~S., {Kenney}, J.~D., {Tacconi}, L., {et~al.} 1986, \apjl, 311, L17,
  \dodoi{10.1086/184790}

\bibitem[{{Zaragoza-Cardiel} {et~al.}(2019){Zaragoza-Cardiel}, {Fritz},
  {Aretxaga}, {Mayya}, {Rosa-Gonz{\'a}lez}, {Beckman}, {Bruzual}, {Charlot}, \&
  {Lomel{\'\i}-N{\'u}{\~n}ez}}]{Zaragoza2019}
{Zaragoza-Cardiel}, J., {Fritz}, J., {Aretxaga}, I., {et~al.} 2019, \mnras,
  487, L61, \dodoi{10.1093/mnrasl/slz093}

\bibitem[{{Zaritsky} \& {Christlein}(2007)}]{Zaritsky2007}
{Zaritsky}, D., \& {Christlein}, D. 2007, \aj, 134, 135, \dodoi{10.1086/518238}

\end{thebibliography}
\bibliographystyle{aasjournal}

\appendix
\section{Supplementry tables}
\begin{table*}
\scriptsize
\centering
\caption{Properties Of FUV bright complexes in NGC 6946. The full table containing all the complexes in all the three galaxies is available in electronic format }
\label{tab:FUV_sfr_radec}
\begin{tabular}{lcc ccc cccr} 
\toprule
Galaxy	&	R.A.(J2000.0)	&	Dec.(J2000.0)	&	Semi-maj.	&	Semi-min.	&	P.A	&	 $\Sigma_{SFR(UV)}$	&  $\Delta\Sigma_{SFR(UV)}$ &	Area	&	Location	\\
(Name)	&	(hh:mm:ss.ss)	&	(dd:mm:ss.ss)	&	(\arcsec)	&	(\arcsec)	&	(degree)	& (M$_\odot$ yr$^{-1}$ kpc$^{-2}$)	& (M$_\odot$ yr$^{-1}$ kpc$^{-2}$)	& (kpc$^{2}$)	&		\\
\hline		 													
NGC 6946	&	20:34:32.23	&	60:04:35.40	&	3.3	&	2.4	&	319	&	0.0047	&	0.0002	&	0.021	&	Inner	\\
NGC 6946	&	20:34:35.65	&	60:04:33.10	&	7.2	&	5.2	&	338	&	0.0139	&	0.0002	&	0.096	&	Inner	\\
NGC 6946	&	20:34:35.84	&	60:04:29.37	&	5.8	&	5.1	&	298	&	0.0178	&	0.0003	&	0.076	&	Inner	\\
NGC 6946	&	20:34:36.24	&	60:04:29.28	&	6.7	&	3.8	&	83	&	0.015	&	0.0003	&	0.065	&	Inner	\\
NGC 6946	&	20:34:36.33	&	60:04:37.78	&	4.6	&	2.9	&	306	&	0.0072	&	0.0002	&	0.035	&	Inner	\\
NGC 6946	&	20:34:20.47	&	59:59:25.16	&	2.2	&	1.7	&	345	&	0.007	&	0.0002	&	0.01	&	Outer	\\
NGC 6946	&	20:34:20.63	&	59:59:27.01	&	2.6	&	1.8	&	320	&	0.006	&	0.0002	&	0.012	&	Outer	\\
NGC 6946	&	20:34:39.81	&	60:02:13.08	&	6.4	&	5.4	&	76	&	0.0064	&	0.0002	&	0.088	&	Outer	\\
NGC 6946	&	20:34:44.35	&	60:02:08.72	&	3.4	&	1.7	&	342	&	0.0046	&	0.0001	&	0.015	&	Outer	\\
NGC 6946	&	20:34:55.33	&	60:01:21.15	&	2	&	1.6	&	302	&	0.0037	&	0.0001	&	0.008	&	Outer	\\
\toprule
\end{tabular}
\end{table*}

\begin{longrotatetable}
\begin{deluxetable}{lcccccc|ccccccr}
\tablecaption{Q value of SFCs in the outer disk \label{table:qvalue}}
\tablewidth{900pt}
\tabletypesize{\scriptsize}
\tablehead{
\colhead{R.A.(J2000.0)} & \colhead{Dec.(J2000.0)} & 
\colhead{Semi-maj} & \colhead{Semi-min} & 
\colhead{P.A} & \colhead{Q value} & \colhead{$\Delta$ Q}
\vline&
\colhead{R.A.(J2000.0)} & \colhead{Dec.(J2000.0)} & 
\colhead{Semi-maj} & \colhead{Semi-min} & 
\colhead{P.A} & \colhead{Q value} & \colhead{$\Delta$ Q} \\
\colhead{(hh:mm:ss.ss)} & \colhead{(dd:mm:ss.ss)} & 
\colhead{(\arcsec)} & \colhead{(\arcsec)} & 
\colhead{(degree)} & \colhead{} & \colhead{}
\vline&
\colhead{(hh:mm:ss.ss)} & \colhead{(dd:mm:ss.ss)} & 
\colhead{(\arcsec)} & \colhead{(\arcsec)} & 
\colhead{(degree)} & \colhead{} & \colhead{}
}
\startdata
&		&	NGC 0628	&	&	&	&	&	&		& NGC 6946		&	&		& &	\\
01:35:59.38	&	15:42:52.92	&	2.8	&	1.9	&	314	&	161.77	&	25.87	&	20:33:28.52	&	60:13:03.27	&	2.3	&	2.2	&	313	&	168.83	&	25.51	\\
01:36:04.25	&	15:45:29.21	&	1.6	&	1.5	&	251	&	1.99	&	0.32	&	20:33:47.66	&	60:07:56.69	&	5.9	&	5.8	&	346	&	1.69	&	0.25	\\
01:36:14.27	&	15:47:33.75	&	2.7	&	1.9	&	262	&	8.58	&	1.37	&	20:33:48.13	&	60:08:21.53	&	3.1	&	2.1	&	50	&	2.15	&	0.33	\\
01:36:16.39	&	15:44:50.68	&	2.9	&	1.5	&	277	&	1.81	&	0.29	&	20:33:48.22	&	60:08:31.79	&	3.3	&	3.0	&	70	&	1.35	&	0.20	\\
01:36:20.03	&	15:44:34.21	&	0.9	&	0.6	&	348	&	4.25	&	0.68	&	20:33:48.72	&	60:08:16.98	&	2.4	&	2.2	&	358	&	1.94	&	0.29	\\
01:36:20.08	&	15:44:36.00	&	1.2	&	0.8	&	277	&	4.06	&	0.65	&	20:33:50.63	&	60:07:04.37	&	3.5	&	2.6	&	334	&	1.93	&	0.29	\\
01:36:23.35	&	15:51:12.32	&	1.2	&	0.7	&	28	&	2.11	&	0.34	&	20:33:58.11	&	60:06:14.62	&	2.3	&	1.8	&	359	&	1.99	&	0.30	\\
01:36:23.35	&	15:42:48.30	&	2.3	&	1.0	&	53	&	4.04	&	0.65	&	20:33:59.42	&	60:05:25.53	&	2.5	&	2.1	&	354	&	3.52	&	0.53	\\
01:36:23.69	&	15:50:58.90	&	3.6	&	2.5	&	315	&	2.07	&	0.33	&	20:34:12.94	&	60:10:13.07	&	2.7	&	2.4	&	289	&	7.45	&	1.13	\\
01:36:25.63	&	15:43:05.15	&	1.8	&	0.9	&	264	&	7.31	&	1.17	&	20:34:20.47	&	59:59:25.16	&	2.2	&	1.7	&	345	&	2.16	&	0.33	\\
01:36:27.55	&	15:51:47.41	&	1.3	&	0.8	&	257	&	2.07	&	0.33	&	20:34:20.63	&	59:59:27.01	&	2.6	&	1.8	&	320	&	2.00	&	0.30	\\
01:36:29.52	&	15:52:05.40	&	2.2	&	2.6	&	360	&	1.63	&	0.26	&	20:34:25.46	&	60:13:13.63	&	2.1	&	1.3	&	14	&	2.49	&	0.38	\\
01:36:34.89	&	15:51:44.41	&	1.8	&	2.1	&	303	&	3.83	&	0.61	&	20:34:28.28	&	60:13:36.22	&	2.8	&	2.2	&	13	&	3.27	&	0.49	\\
01:36:38.98	&	15:53:08.94	&	1.5	&	0.6	&	305	&	2.51	&	0.40	&	20:34:29.13	&	60:13:48.71	&	5.1	&	4.0	&	76	&	2.94	&	0.44	\\
01:36:42.48	&	15:52:50.86	&	2.0	&	2.2	&	360	&	1.15	&	0.18	&	20:34:30.30	&	60:04:16.87	&	2.6	&	1.7	&	78	&	2.92	&	0.44	\\
01:36:43.13	&	15:52:43.92	&	1.7	&	0.7	&	18	&	1.31	&	0.21	&	20:34:39.81	&	60:02:13.08	&	6.4	&	5.4	&	76	&	3.43	&	0.52	\\
01:36:45.74	&	15:52:59.78	&	0.9	&	0.6	&	349	&	1.50	&	0.24	&	20:34:43.84	&	60:15:07.09	&	1.9	&	1.5	&	314	&	2.42	&	0.37	\\
01:36:48.75	&	15:41:09.67	&	3.0	&	2.6	&	36	&	1.40	&	0.22	&	20:34:44.35	&	60:02:08.72	&	3.4	&	1.7	&	342	&	3.69	&	0.56	\\
01:36:49.29	&	15:53:03.26	&	1.0	&	0.6	&	327	&	1.53	&	0.24	&	20:34:46.65	&	60:03:42.72	&	3.0	&	1.9	&	31	&	3.51	&	0.53	\\
01:36:49.39	&	15:41:45.22	&	2.7	&	1.8	&	312	&	3.69	&	0.59	&	20:34:46.69	&	60:02:35.03	&	2.5	&	1.4	&	341	&	7.52	&	1.14	\\
01:36:50.42	&	15:53:09.79	&	1.7	&	0.6	&	354	&	1.53	&	0.24	&	20:34:46.77	&	60:02:37.70	&	3.0	&	1.2	&	313	&	7.88	&	1.19	\\
01:36:51.63	&	15:41:36.39	&	4.0	&	3.1	&	276	&	1.20	&	0.19	&	20:34:47.03	&	60:03:33.59	&	2.8	&	2.7	&	314	&	2.21	&	0.33	\\
01:36:53.85	&	15:42:07.70	&	2.3	&	2.0	&	335	&	1.59	&	0.25	&	20:34:47.36	&	60:02:36.31	&	2.0	&	1.9	&	27	&	6.61	&	1.00	\\
01:36:53.86	&	15:41:48.67	&	2.7	&	1.2	&	261	&	2.34	&	0.37	&	20:34:47.64	&	60:02:44.06	&	2.3	&	1.6	&	280	&	3.52	&	0.53	\\
01:36:53.91	&	15:53:12.56	&	5.2	&	1.2	&	20	&	1.65	&	0.26	&	20:34:48.67	&	60:02:36.57	&	4.1	&	2.3	&	312	&	2.62	&	0.40	\\
01:36:54.02	&	15:53:13.36	&	3.1	&	1.3	&	275	&	1.60	&	0.26	&	20:34:51.32	&	60:14:20.58	&	3.7	&	3.3	&	278	&	46.53	&	7.03	\\
01:36:54.93	&	15:52:57.85	&	4.4	&	3.0	&	323	&	2.88	&	0.46	&	20:34:54.03	&	60:15:43.84	&	2.2	&	1.4	&	1	&	2.13	&	0.32	\\
01:37:00.49	&	15:51:20.10	&	2.0	&	1.6	&	277	&	1.88	&	0.30	&	20:34:55.33	&	60:01:21.15	&	2.0	&	1.6	&	302	&	15.94	&	2.41	\\
01:37:01.89	&	15:45:53.70	&	1.8	&	0.7	&	279	&	1.14	&	0.18	&	20:34:56.34	&	60:15:35.96	&	3.9	&	2.9	&	342	&	5.13	&	0.78	\\
01:37:02.05	&	15:54:46.70	&	1.6	&	1.4	&	310	&	14.97	&	2.39	&	20:35:03.70	&	60:15:49.29	&	2.4	&	1.4	&	86	&	3.23	&	0.49	\\
01:37:02.32	&	15:45:01.84	&	1.4	&	0.7	&	262	&	2.30	&	0.37	&	20:35:11.87	&	60:04:02.48	&	1.8	&	1.5	&	43	&	5.31	&	0.80	\\
01:37:02.41	&	15:46:48.62	&	1.6	&	0.9	&	272	&	1.23	&	0.20	&	20:35:15.30	&	60:16:32.75	&	2.2	&	2.1	&	35	&	1.92	&	0.29	\\
01:37:02.60	&	15:47:11.80	&	3.2	&	2.3	&	262	&	2.03	&	0.32	&	20:35:15.57	&	60:05:16.71	&	2.3	&	1.9	&	280	&	3.41	&	0.52	\\
01:37:03.16	&	15:47:02.36	&	2.5	&	0.9	&	269	&	1.76	&	0.28	&	20:35:17.26	&	60:16:33.90	&	1.7	&	1.6	&	284	&	3.05	&	0.46	\\
01:37:03.29	&	15:47:51.12	&	2.4	&	1.8	&	260	&	2.00	&	0.32	&	20:35:17.33	&	60:16:41.85	&	2.5	&	2.1	&	346	&	5.47	&	0.83	\\
01:37:04.55	&	15:48:00.37	&	3.9	&	3.2	&	42	&	2.68	&	0.43	&	20:35:17.92	&	60:05:19.36	&	2.9	&	2.1	&	286	&	1.73	&	0.26	\\
&		&	NGC 0628	&	&	&	&	&	&		& NGC 6946		&	&		& &	\\
01:37:05.34	&	15:44:46.27	&	2.4	&	1.3	&	285	&	4.51	&	0.72	&	20:35:19.58	&	60:15:23.08	&	3.6	&	3.0	&	89	&	12.92	&	1.95	\\
01:37:07.21	&	15:39:29.83	&	1.6	&	0.8	&	251	&	5.64	&	0.90	&	20:35:21.81	&	60:04:12.15	&	2.8	&	2.5	&	324	&	2.33	&	0.35	\\
01:37:16.02	&	15:50:31.79	&	3.7	&	2.7	&	289	&	6.14	&	0.98	&	20:35:31.27	&	60:04:39.69	&	2.2	&	1.5	&	297	&	4.19	&	0.63	\\
 	&	 	&	NGC 5457	&	 	&	 	&	 	&		&	20:35:32.08	&	60:07:22.07	&	2.2	&	2.2	&	353	&	2.78	&	0.42	\\
14:02:10.64	&	54:30:10.37	&	1.3	&	0.9	&	252	&	0.64	&	0.10	&	20:35:32.11	&	60:16:33.91	&	2.9	&	2.5	&	300	&	2.60	&	0.39	\\
14:02:16.76	&	54:30:09.74	&	2.1	&	1.6	&	310	&	0.34	&	0.05	&	20:35:32.33	&	60:07:21.63	&	2.2	&	1.8	&	29	&	2.65	&	0.40	\\
14:02:45.79	&	54:33:43.81	&	1.9	&	1.2	&	177	&	0.21	&	0.03	&	20:35:32.77	&	60:15:14.89	&	2.0	&	1.9	&	60	&	248.36	&	37.52	\\
14:02:47.93	&	54:33:54.49	&	1.9	&	1.7	&	312	&	0.22	&	0.03	&	20:35:33.87	&	60:05:03.58	&	2.6	&	2.3	&	300	&	2.39	&	0.36	\\
14:03:10.51	&	54:35:52.41	&	1.7	&	1.3	&	188	&	0.11	&	0.02	&	20:35:34.11	&	60:05:22.66	&	3.6	&	3.5	&	303	&	4.85	&	0.73	\\
14:03:11.94	&	54:35:41.53	&	7.8	&	3.9	&	228	&	0.07	&	0.01	&	20:35:35.15	&	60:08:49.88	&	1.6	&	1.6	&	359	&	5.31	&	0.80	\\
14:03:12.52	&	54:35:46.93	&	5.2	&	4.2	&	196	&	0.07	&	0.01	&	20:35:35.44	&	60:10:19.13	&	2.4	&	1.8	&	58	&	4.40	&	0.66	\\
14:03:13.44	&	54:35:41.28	&	12.9	&	10.0	&	178	&	0.08	&	0.01	&	20:35:36.49	&	60:08:32.44	&	2.1	&	1.8	&	35	&	5.27	&	0.80	\\
14:03:18.65	&	54:35:21.49	&	5.8	&	3.2	&	297	&	0.13	&	0.02	&	20:35:36.57	&	60:11:04.55	&	2.4	&	1.7	&	44	&	2.51	&	0.38	\\
14:03:32.08	&	54:32:59.94	&	3.3	&	2.7	&	324	&	0.22	&	0.03	&	20:35:36.84	&	60:17:14.78	&	3.1	&	2.5	&	87	&	3.29	&	0.50	\\
14:03:32.34	&	54:32:57.66	&	2.0	&	1.6	&	169	&	0.18	&	0.03	&	20:35:37.34	&	60:05:00.65	&	1.8	&	1.4	&	271	&	2.55	&	0.38	\\
14:03:35.33	&	54:33:17.76	&	1.0	&	0.6	&	234	&	0.49	&	0.07	&	20:35:37.45	&	60:06:59.80	&	2.4	&	1.9	&	87	&	3.19	&	0.48	\\
14:03:37.05	&	54:32:53.61	&	4.2	&	2.6	&	239	&	0.52	&	0.08	&	20:35:37.62	&	60:05:04.13	&	2.0	&	1.3	&	19	&	2.78	&	0.42	\\
14:04:23.66	&	54:32:58.57	&	0.7	&	0.7	&	190	&	0.80	&	0.12	&	20:35:42.59	&	60:12:03.11	&	1.7	&	1.7	&	358	&	3.58	&	0.54	\\
14:04:29.48	&	54:25:44.27	&	2.4	&	1.5	&	152	&	0.14	&	0.02	&	20:35:43.82	&	60:11:49.36	&	2.5	&	2.4	&	328	&	3.17	&	0.48	\\
14:04:31.95	&	54:25:26.00	&	4.0	&	3.3	&	200	&	0.15	&	0.02	&	20:35:46.40	&	60:08:24.79	&	3.9	&	3.3	&	310	&	2.16	&	0.33	\\
14:04:32.02	&	54:26:31.19	&	2.0	&	1.4	&	200	&	0.21	&	0.03	&	20:35:46.43	&	60:08:22.93	&	3.6	&	2.1	&	337	&	2.50	&	0.38	\\
14:04:35.00	&	54:27:00.15	&	1.0	&	0.5	&	163	&	0.25	&	0.04	&	20:35:46.56	&	60:08:29.15	&	4.7	&	4.4	&	318	&	1.91	&	0.29	\\
14:04:35.11	&	54:23:12.02	&	0.9	&	0.6	&	217	&	0.22	&	0.03	&	20:35:46.74	&	60:17:32.58	&	2.7	&	1.4	&	50	&	1.88	&	0.28	\\
14:04:35.60	&	54:20:22.33	&	2.5	&	1.2	&	167	&	0.31	&	0.05	&	20:35:46.81	&	60:08:30.98	&	5.8	&	3.6	&	320	&	1.91	&	0.29	\\
14:04:35.69	&	54:20:18.65	&	2.3	&	0.9	&	211	&	0.43	&	0.06	&	20:35:46.91	&	60:08:35.93	&	3.1	&	2.5	&	25	&	1.63	&	0.25	\\
14:04:36.16	&	54:25:01.17	&	4.0	&	2.0	&	312	&	0.23	&	0.03	&	20:35:47.37	&	60:08:59.14	&	2.6	&	1.7	&	329	&	6.35	&	0.96	\\
14:04:36.41	&	54:23:15.44	&	2.5	&	2.2	&	320	&	0.25	&	0.04	&	20:35:53.15	&	60:14:38.90	&	2.2	&	1.6	&	37	&	3.48	&	0.53	\\
14:04:36.86	&	54:23:10.53	&	1.7	&	1.0	&	280	&	0.25	&	0.04	&	20:35:56.58	&	60:03:22.25	&	2.1	&	1.3	&	5	&	7.29	&	1.10	\\
14:04:37.08	&	54:26:55.52	&	3.0	&	1.7	&	146	&	0.15	&	0.02	&		&		&		&		&		&		&		\\
14:04:37.16	&	54:27:04.72	&	1.0	&	0.6	&	218	&	0.15	&	0.02	&		&		&		&		&		&		&		\\
14:04:37.53	&	54:20:29.68	&	1.9	&	1.3	&	238	&	0.27	&	0.04	&		&		&		&		&		&		&		\\
14:04:38.22	&	54:23:24.20	&	2.6	&	2.1	&	219	&	0.26	&	0.04	&		&		&		&		&		&		&		\\
14:04:38.46	&	54:23:48.09	&	3.6	&	2.7	&	212	&	0.25	&	0.04	&		&		&		&		&		&		&		\\
14:04:39.02	&	54:21:36.50	&	2.1	&	0.9	&	277	&	0.26	&	0.04	&		&		&		&		&		&		&		\\
14:04:39.86	&	54:23:55.95	&	3.6	&	2.2	&	210	&	0.31	&	0.05	&		&		&		&		&		&		&		\\
14:04:40.00	&	54:15:12.17	&	1.4	&	0.5	&	205	&	1.07	&	0.16	&		&		&		&		&		&		&		\\
14:04:40.11	&	54:23:58.97	&	2.4	&	1.1	&	173	&	0.34	&	0.05	&		&		&		&		&		&		&		\\
14:04:40.52	&	54:23:52.46	&	3.2	&	2.4	&	272	&	0.39	&	0.06	&		&		&		&		&		&		&		\\
&		&	NGC 5457	&	&	&	&	&	&		& 		&	&		& &	\\
14:04:40.91	&	54:23:51.14	&	2.8	&	1.4	&	236	&	0.50	&	0.08	&		&		&		&		&		&		&		\\
14:04:41.03	&	54:23:55.39	&	1.9	&	0.9	&	265	&	0.39	&	0.06	&		&		&		&		&		&		&		\\
14:04:49.03	&	54:22:49.46	&	1.7	&	1.2	&	245	&	0.19	&	0.03	&		&		&		&		&		&		&		\\
14:04:49.63	&	54:28:17.93	&	1.3	&	1.0	&	301	&	0.41	&	0.06	&		&		&		&		&		&		&		\\
14:04:50.63	&	54:28:07.45	&	3.9	&	2.8	&	149	&	0.49	&	0.07	&		&		&		&		&		&		&		\\
\enddata
\end{deluxetable}
\end{longrotatetable}

\end{document}